\documentclass[12pt,preprint]{aastex}
\newcommand{\Lya}{\mbox{Ly\,{\sc $\alpha$}}}
\newcommand{\nion}{$N_{\rm ion}$}
\newcommand{\hst}{{\it HST}}
\newcommand{\kms}{km~s$^{-1}$}
\newcommand{\cm}{cm$^{-2}$}
\newcommand{\cmth}{cm$^{-3}$}
\newcommand{\ngc}{NGC~4395}
\newcommand{\Ah}{A$_{\rm h}$}
\newcommand{\Al}{A$_{\rm l}$}

\shorttitle{MULTIWAVELENGTH MONITORING OF NGC~4395. IV.} \shortauthors{BASKIN \& LAOR}
\begin{document}
\title{Multiwavelength Monitoring of the Dwarf Seyfert 1 Galaxy NGC~4395. IV. The Variable UV Absorption Lines}
\author{Alexei Baskin and Ari Laor}
\affil{Physics Department, Technion, Haifa~32000, Israel}
\email{alexei@physics.technion.ac.il; laor@physics.technion.ac.il}

\begin{abstract}
We report the detection of variable UV absorption lines in \ngc, based on UV observations with the \hst\ STIS carried out in April and July, 2004, as part of a reverberation-mapping campaign. Low-ionization lines of \ion{O}{1}, \ion{N}{1}, \ion{Si}{2}, \ion{C}{2}, and \ion{Fe}{2}, are present in the low-state spectra (April 2004) at a velocity $v_{\rm shift}=-250$~\kms\ (system \Al), and additional high-ionization lines of \ion{C}{4} and \ion{N}{5} appear in the high-state spectra (July 2004) at $v_{\rm shift}=-250$~\kms\ (system \Ah) and at $v_{\rm shift}=-840$~\kms\ (system B). The absence of absorption from the low metastable levels of \ion{Si}{2} implies a density $\la 10^3$~cm$^{-3}$ for system \Al, indicating a location outside the narrow line region (NLR). System \Ah\ is peculiar as only \ion{N}{5} absorption is clearly detected. A high \ion{N}{5}/\ion{C}{4} absorption ratio is expected for a high metallicity absorber, but this is excluded here as the metallicity of the host galaxy and of the nuclear gas is significantly subsolar. A simple acceptable model for systems \Ah\ and B is an absorber located between the broad line region (BLR) and the NLR, which absorbs only the continuum and the BLR. At the low-state the strong narrow emission lines of \ion{C}{4} and \ion{N}{5} dominate the spectrum, making the absorption invisible. At the high-state the absorbed continuum and BLR emission dominate the spectrum. Thus, the change in the observed absorption does not reflect a change in the absorber, but rather a change in the continuum and BLR emission from behind the absorber, relative to the emission from the NLR in front of the absorber. Studies of the absorption line variability in highly variable objects can thus break the degeneracy in the absorber distance determination inherent to single epoch studies.
\end{abstract}

\section{INTRODUCTION}
The Sd III--IV dwarf galaxy \ngc\ harbors a Seyfert 1 nucleus with $L_{\rm bol} \sim 10^{40}$~erg~s$^{-1}$, by far the lowest luminosity type 1 active galactic nucleus (AGN; Filippenko \& Sargent 1989; Filippenko, Ho, \& Sargent 1993; Filippenko \& Ho 2003). A recent reverberation-mapping campaign yielded a black hole mass of $M_{\bullet}=(3.6\pm1.1)\times10^5 M_{\sun}$ (Peterson et~al.\ 2005), among the lowest values measured in AGN. Thus, \ngc\ offers the opportunity to explore AGN at the lowest end of the mass and luminosity scales. In particular, recent studies of its emission line properties provide interesting constraints on the nature of the broad line region (BLR, Laor 2007 and references therein). In this paper we study the UV absorption lines in this object, and their time variability, as revealed in the reverberation-mapping campaign spectra.

About $\sim50$\% of Seyfert 1 galaxies show outflows of ionized gas from their nuclei (Crenshaw et~al.\ 1999), which is manifested by intrinsic UV and X-ray absorption-lines that are blue-shifted with respect to the systemic velocities of the galaxies. At luminosities higher than those of Seyferts, a significant fraction of quasars (25--60\%) shows intrinsic absorption in the UV (Ganguly et~al.\ 2001; Laor \& Brandt 2002; Vestergaard 2003; Ganguly \& Brotherton 2008) and in the X-ray (George et~al. 2000). The lowest luminosity and lowest black hole mass object where absorption was detected until now is NGC~4051, which has $L_{\rm bol} = 2.7\times10^{43}$~ergs~s$^{-1}$ (Ogle et~al.\ 2004) and $M_{\bullet}\sim1.9\times10^6 M_{\sun}$ (Peterson et~al.\ 2004). It is a well studied object with reported short-term continuum flux variability in the X-ray, UV, and optical bands (e.g.\ McHardy et~al.\ 2003; Uttley et~al.\ 2000; Shemmer et~al.\ 2003; respectively). Intrinsic absorption systems were detected in the X-ray and UV spectra of NGC~4051 (Kaspi et~al.\ 2004, and references therein).

Earlier UV spectroscopy of \ngc\ with the \hst\ FOS (Kraemer et~al.\ 1999) and at a higher spectral resolution (but lower S/N) with the \hst\ STIS echelle (Crenshaw et~al.\ 2004) revealed absorption from ionized gas in the line of sight to the nucleus in two systems. One with an outflow velocity of $v=-770$~\kms\ (at rest frame), detected only in \ion{C}{4}, and the second at $v=-114$~\kms, detected in \ion{C}{4}, \ion{Si}{4}, and at lower ionization states in \ion{Mg}{2}, \ion{O}{1}, and \ion{C}{2}. The first system is attributed to an ionized outflowing wind, while the second is attributed to absorption by the ISM of \ngc, however these conclusions were not robust due to the low S/N of the STIS echelle spectra.

The large variability of \ngc\ in the X-ray, UV and optical bands (Lira et~al.\ 1999; Moran et~al.\ 1999, 2005; Iwasawa et~al.\ 2000; Shih, Iwasawa, \& Fabian 2003; Vaughan et~al.\ 2005; Peterson et~al.\ 2005; O'Neill et~al.\ 2006; Desroches et~al.\ 2006) provides an opportunity to distinguish between absorption by foreground gas far from the nucleus, which is likely to be non-variable, and absorption by gas intrinsic to the nucleus, which may be variable. The nature of the variability can provide constraints on the geometry and physical properties of the absorber.

We adopt a modification of the commonly used absorption-line measurement techniques (see a review by Crenshaw, Kraemer, \& George 2003, \S~2 there) to measure the absorption-line properties, i.e.\ the ionic column-density, the covering factor, and the intrinsic line width. The adopted technique should allow a more realistic estimate of the possible range of the absorption-line properties. We generate a grid of synthetic profiles per an absorption line based on atomic physics only, thus avoiding the theoretical biases commonly introduced by deducing the ionic column-densities based on an assumed photoionization model (e.g., Laor et~al.\ 1997). The {\it acceptable} range of the absorption-line properties is then defined by the range of the synthetic absorption-line profiles which produce an ``acceptable'' reconstruction of the observed spectrum, which is produced by dividing the observed spectrum by the synthetic profile. This allows a more realistic estimate of the possible uncertainties than estimating the observed absorption-line profile using a fit of the unabsorbed spectrum that is based on an interpolated spectral shape. Only after estimating the observationally acceptable range of the absorption-line properties, we search for the possible photoionization model solutions.

In this paper, we describe the results of an effort to confirm the presence of intrinsic UV absorption towards the nucleus of \ngc\ and determine its properties. The study is based on \hst\ STIS spectra obtained as part of a reverberation-mapping program on \ngc, which revealed large continuum variability (by a factor of 4--7), as well as large changes in the absorption spectrum. The observations and data processing are outlined in \S~2. The absorption-line measurement process is outlined in \S~3, and the photoionization models used to estimate the absorber properties are described in \S~4. A discussion of the origin of the absorber and its properties, and a comparison to earlier studies are presented in \S~5. Our conclusions are summarized in \S~6.

\section{UV OBSERVATIONS}
\subsection{Data acquisition and processing}
A complete description of the observations and data reduction is provided in Peterson et~al.\ (2005), and is briefly reviewed here. \ngc\ was observed with the \hst\ Space Telescope Imaging Spectrograph (STIS) using the FUV MAMA detector with the G140L grating, which covers the spectral range 1150--1700~\AA\ at 0.5831~\AA\ per pixel and a velocity resolution of $\sim$~150--250~\kms~FWHM. Three sessions of observations took place: the first session, visit 1, began on 2004 April 10; the second session, visit 2, began on 2004 April 11; and a third later session, visit 3, began on 2004 July 3 (all dates are UT dates). Each visit is composed of five orbits.  The flux density uncertainties are derived using the standard-deviation of the flux density in the (presumably featureless) 1340--1360~\AA\ continuum window, scaled to other wavelengths assuming photon statistics.

The photometric accuracy of the observation during visit 1 was degraded as a result of a slow drift of the target in the aperture (Peterson et~al.\ 2005), which did not allow to use these spectra for reverberation mappings. These spectra are still useful for our purpose of studying the absorption-line profiles, as an accurate photometry is not required. However, the drift also resulted in a shift of the spectra along the wavelength scale which we had to correct for. The following procedure was executed to align in wavelength-scale the spectra of the five orbits composing visit 1. Each spectrum was cross-correlated with the first orbit spectrum using a wavelength-shift ($\Delta\lambda$) as a free parameter, where we searched for the wavelength-shift which gives the maximum cross-correlation i.e., $\max_{(\Delta\lambda)_i}\int f_1[\lambda]\cdot f_i\left[\lambda+(\Delta\lambda)_i\right]d\lambda$, where $f_i$ is the flux of the $i$-th orbit spectrum, $i=2$--5. Since the continuum does not contribute to the cross-correlation signal, we integrated only over wavelength regions containing the strong emission features of \Lya~$\lambda1215.67$, \ion{C}{4}~$\lambda1549.05$ and \ion{He}{2}~$\lambda1640.42$. The resulting shifts for orbits 2, 3, 4 and 5 were 0.56, 0.88, 1.36, and 1.95 pixels, respectively. The five orbits were then co-added to form the mean spectrum of visit 1, and the shift of the mean spectrum was derived by cross-correlating it with the mean spectrum of visit 2, yielding a wavelength-shift of $\Delta\lambda=-0.89$~pixel ($\sim-0.52$~\AA).

An eye inspection suggested that the absorption features did not vary during the span of the 5 orbits, for each of the visits. To investigate this quantitatively we calculated the $\chi^2$ between all orbits and orbit 3, for each of the visits. The spectrum of each orbit was first normalized locally in the $\pm1500$~\kms\ range of each absorption-line feature (listed in \S~2.2) by dividing the local spectrum by the total flux at that range. Each orbit was then compared to orbit 3 by calculating the $\chi^2_{\rm local}$ value for the prominent absorption feature, using the locally-normalized spectra, and then summing over the individual values of $\chi^2_{\rm local}$. We used the \ion{C}{2}, \ion{N}{1}, \ion{O}{1}, and \ion{Si}{2} absorption lines for all visits, with the addition of \ion{N}{5} in visit 3. This procedure yielded $\chi^2$/degrees of freedom (dof) values in the range of 112--195/130 for visit 1, 135--210/130 for visit 2, and 183--270/170 for visit 3. The $\chi^2/$dof values are thus generally not well above 1, and since they are most likely affected by some small systematic noise, in addition to the pure statistical noise, we conclude that there is no strong evidence for variability of the absorption-line profiles on an orbit timescale. We then co-added the spectra of the individual five orbits for each visit, to form the average spectrum of each visit, which we analyze below. Finally, a red-shift of $z=0.00106$ was used to transform the observed wavelength-scale to the rest-frame of \ngc, based on 21~cm \ion{H}{1} line measurements (Haynes et~al.\ 1998; Springob et~al.\ 2005). The spectra were corrected for Galactic extinction using $E(B-V)=0.017$~mag.\ from Schlegel, Finkbeiner, \& Davis (1998, as listed in the NASA/IPAC Extragalactic Database) and the reddening law of Seaton (1979).

\subsection{Comparison between the three visits}
Figure 1 [panels (a1,a2)] presents the average spectra of the three visits. \ngc\ was in a low-flux state during visits 1 and 2, which we denote by ``low 1'' and ``low 2'', and in a high-flux state during visit 3, which we denote by ``high''.  The rest-fame positions of various absorption lines are marked above the spectra. A possible \ion{C}{1} + \ion{Si}{1} absorption blend at 1560~\AA\ is observed only in the low-state spectra, but as we discuss below this feature may not be real. A question mark in the low-states near 1524~\AA\ marks an additional possible absorption feature. The wavelength range 1207.7--1221.7~\AA\ was significantly contaminated by the Geocoronal \Lya\ emission, which increased the count-rate per pixel by a factor of twenty in the raw image data. Although the \hst\ pipeline reduction is designed to subtract the background Geocoronal line contribution, the corrected spectrum fails to show the damped Galactic \Lya\ absorption line, which must be very prominent. This indicates that the Geocoronal contamination was too strong to correct for accurately, and we therefore exclude the 1207.7--1221.7~\AA\ region, marked by grayed area in Fig.\ 1, from further analysis. The region above $\sim1650$~\AA\ was also excluded from the analysis because of the poor continuum S/N of the low-state spectra. Panels (b1,b2) of Figure 1 present the low 2/low 1 flux ratio. Broad features are present near the major emission lines, indicating differences in the variability levels of the lines and the continuum. However, no clear features are present in the ratio plot near the absorption features, indicating that the absorption profiles are not significantly variable on a one day timescale (the only exception is the \ion{C}{1} + \ion{Si}{1} absorption blend, which is not clearly present in the low 1 state). We therefore co-added the low 1 and low 2 spectra, which we denote as the ``low'' state spectrum, and in panels (c1,c2) we show the high/low flux density ratio. Again, prominent features are generally not present near the low-ionization absorption lines of \ion{C}{2}, \ion{N}{1}, \ion{O}{1}, and \ion{Si}{2}, suggesting they are not variable. A very prominent feature is present near the high-ionization \ion{N}{5} line, where the absorption is strong in the high-state, and is absent in the low-state. Note also the large overall continuum increase in the high-state, from a factor of $\sim 4$ near 1700~\AA, to $\sim 7$ near 1150~\AA. Panels (d1,d2) show the flux difference between the high- and low-states. All absorption lines are clearly seen, indicating that the added flux in the high-state went through both the low- and high-ionization ions. Figures 2 and 3 present a close-up of the low- and high-ionization absorption features, respectively. A prominent absorption feature at $v_{\rm shift}=-250$~\kms\ can be discerned for most of the emission features in Figures 2 and 3, which we denote as systems \Al\ and \Ah\ for the low- and high-ionization absorbers, respectively. An absorption feature at $v_{\rm shift}=-840$~\kms\ is seen in the high-state in \ion{C}{4}, which we denote as system B. The measured absorption equivalent width (EW) are presented in Table 1 (under ${\rm CF}=1$, where CF is the covering factor), and are typically at a level of 1~\AA, or lower.
\placefigure{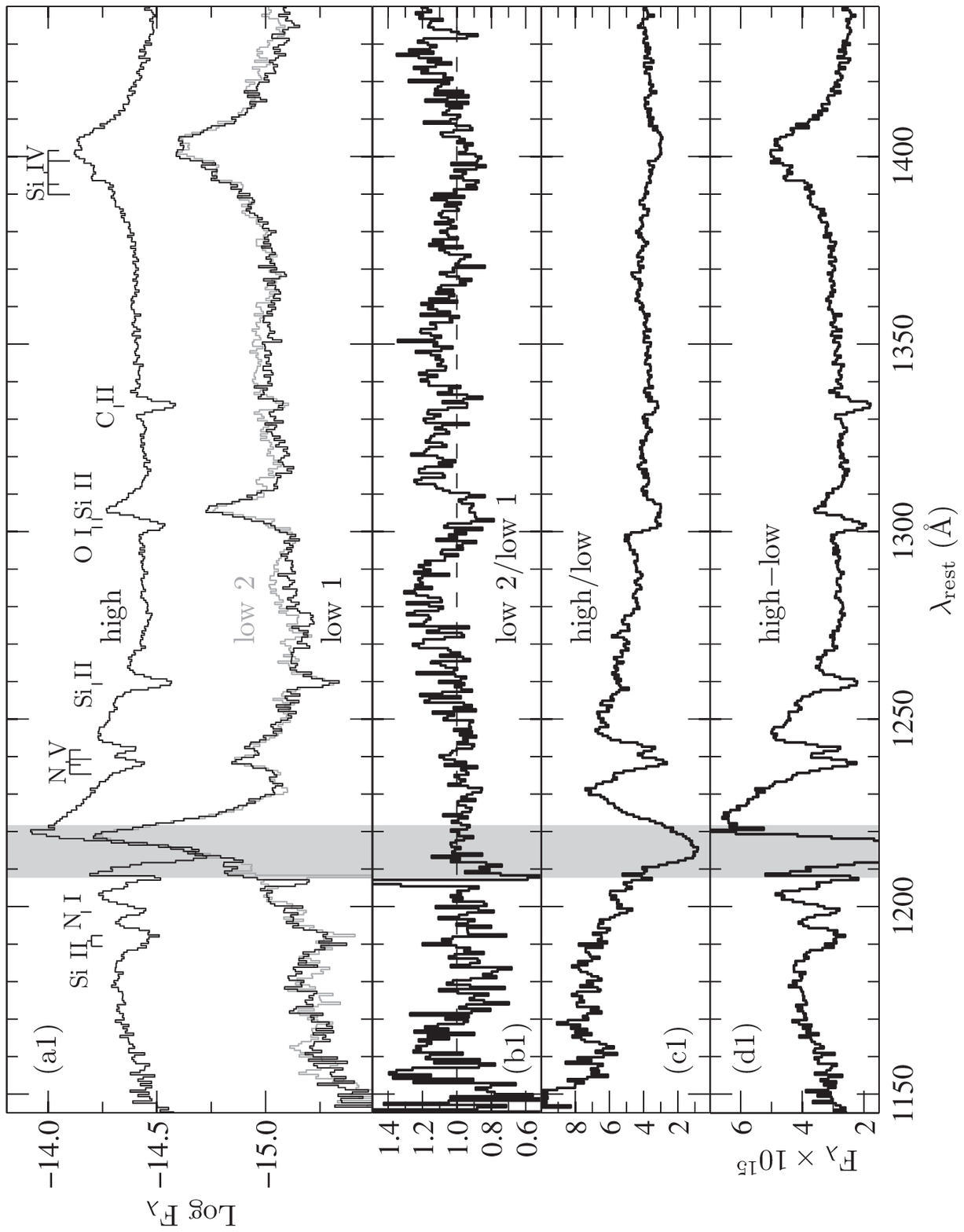,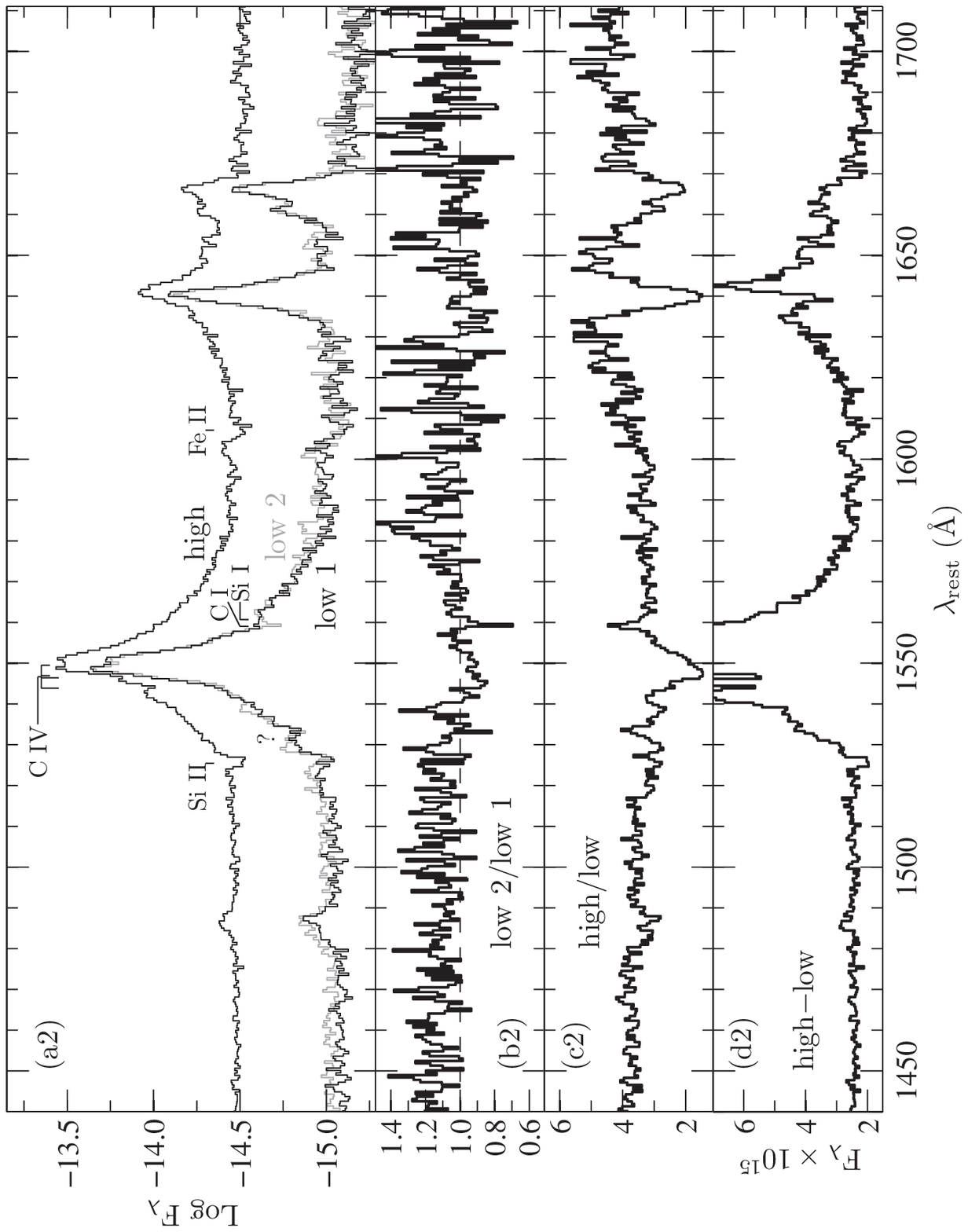} \placefigure{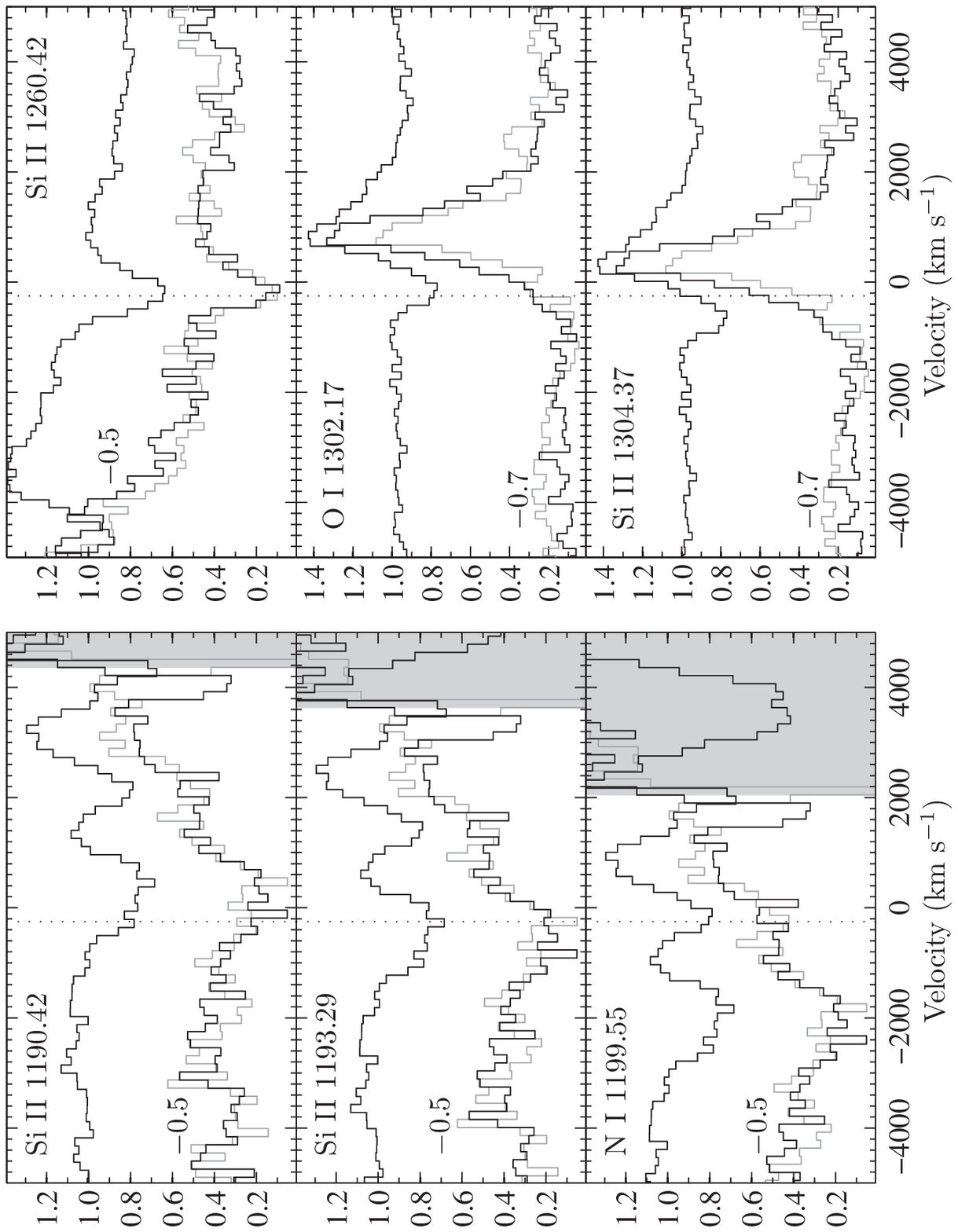,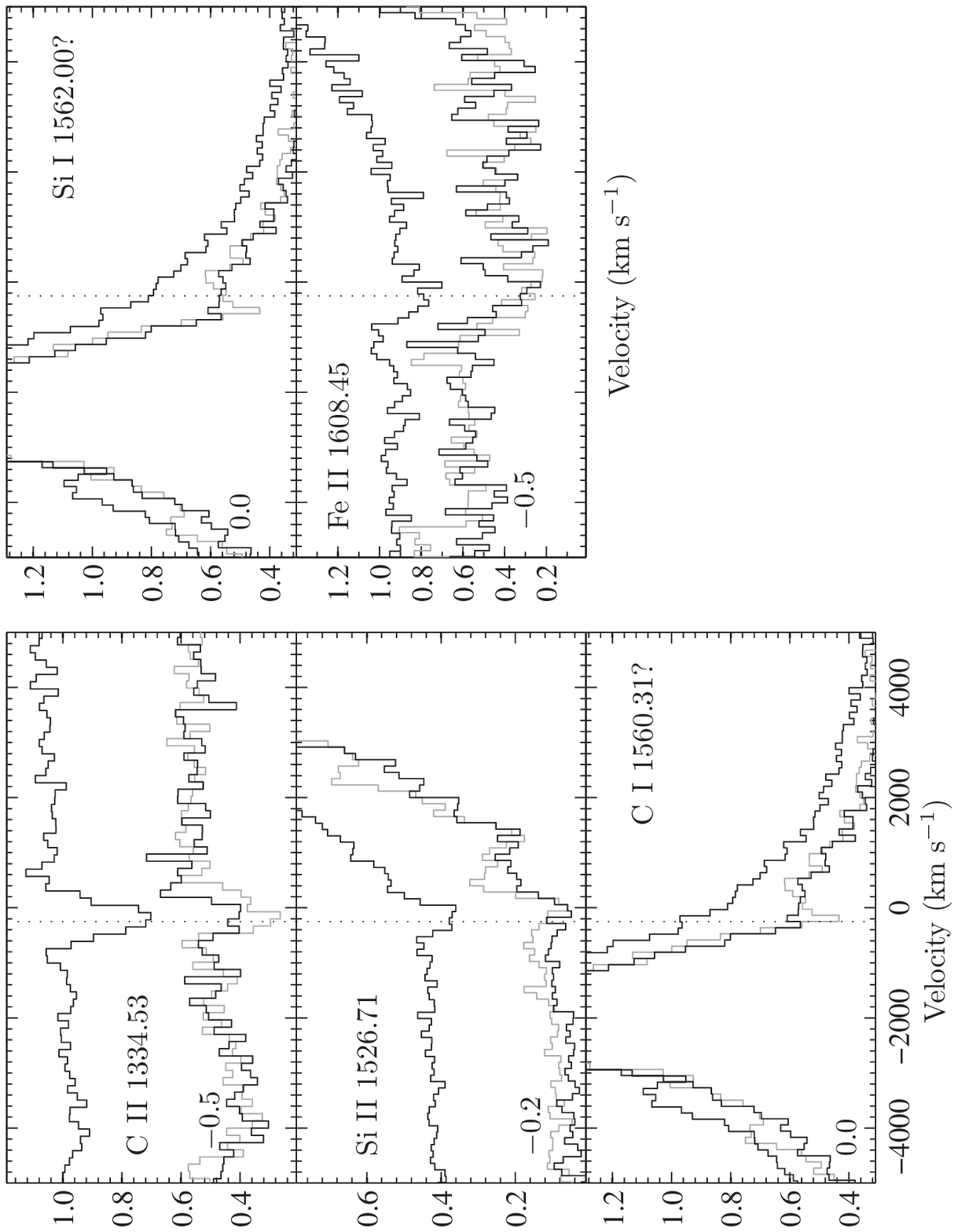} \placefigure{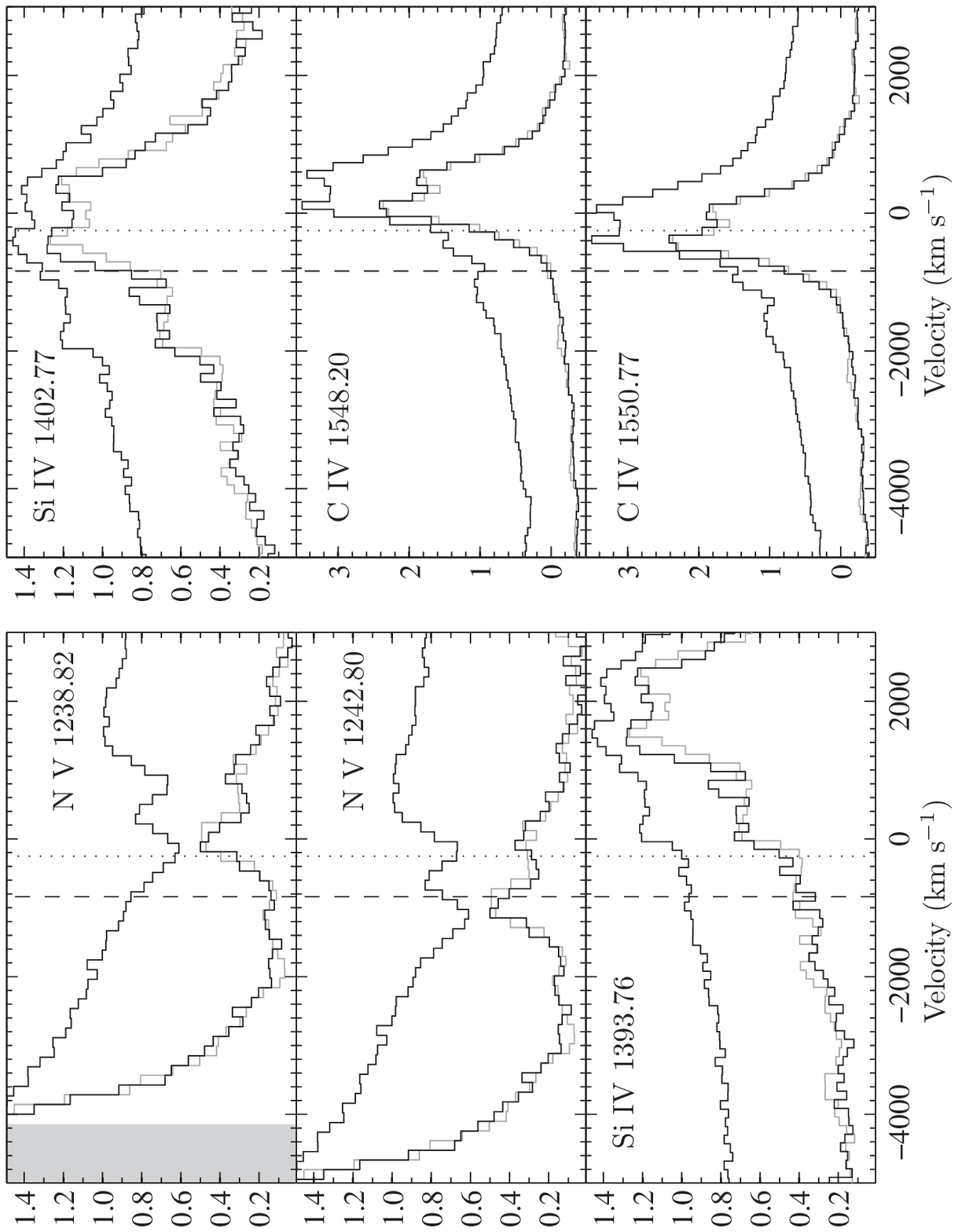}

\section{ABSORPTION-LINE PROFILE RECONSTRUCTION}
We present below a brief review of the technique we use to measure the UV absorption profile, the motivation to use it, and the physical mechanisms incorporated by this technique.

\subsection{Physical broadening mechanisms}
First, we discuss the mechanisms which control the shape of an absorption-line profile as a function of the radial velocity. This shape is set by a convolution of the natural Lorentzian line profile with the thermal velocity distribution, which yields the Voigt profile, $\Psi(v)$ (e.g.\ Rybicki \& Lightman 1979, Eq.~10.77 there). This profile is then convolved with the bulk radial-velocity distribution of the absorbing ionic column, $N_{\rm ion}(v)$. We write $N_{\rm ion}(v)=N_{\rm ion}\cdot f_{\rm ion}(v)$, where \nion\ is the total ionic column-density of the gas and $f_{\rm ion}(v)$ is the normalized distribution of the ionic column-density i.e., $\int_{-\infty}^{\infty}f_{\rm ion}(v)dv=1$. Since the absorption profiles we measure here are generally unresolved, we parameterize $f_{\rm ion}(v)$ as a Gaussian. The convolution of the bulk radial velocity distribution and the thermal one leads to a Gaussian velocity distribution, $f(v)\sim e^{-v^2/(2b^2)}$, with $b^2=b_{\rm thermal}^2+b_{\rm bulk}^2$. A third mechanism that controls the shape of the absorption-line profile is the fraction of the illuminating source covered by the absorbing gas at a given radial velocity, ${\rm CF}(v)$.

In a curve-of-growth analysis the implied $N_{\rm ion}$ for a given absorption EW is a function of $b$. The minimum possible value, $b=b_{\rm thermal}\le 10$~\kms, obtained for $b_{\rm bulk}=0$~\kms, leads to an upper limit on $N_{\rm ion}$. The other extreme, $b=b_{\rm bulk}\gg b_{\rm thermal}$ leads to a lower limit on $N_{\rm ion}$. In the first case the intrinsic absorption is generally highly saturated, while in the second it is generally optically thin.

We use a {\sc FORTRAN77} subroutine {\sc HUMLIK} (Wells 1999) to calculate the Voigt function. The normalized intrinsic absorption profile that includes the effects of the line-of-sight CF and optical depth ($\tau$) equals
\begin{equation}
f_{\rm abs}(v)=[1-{\rm CF}(v)]+{\rm CF}(v)\cdot e^{-\tau(v)}.\label{I}
\end{equation}
Thus, to model a given intrinsic absorption-line profile one should find four independent parameters: $N_{\rm ion}$, $b$, CF$(v)$ (at each velocity segment $v$), and $v_{\rm shift}$ -- the overall velocity-shift of the absorption system relative to the systemic red-shift of the object.

\subsection{Instrumental broadening}
The absorption profiles we detect in \ngc\ generally appear to be unresolved, or possibly marginally resolved. Here we make some simulations to better understand how the \hst\ STIS instrumental resolution affects lines which are marginally resolved. The FWHM of the \hst\ STIS line-spread function (LSF) is $\sim 150-200$~\kms, and one expects that broader lines will not be significantly distorted by the LSF. However, the STIS LSF has a strong narrow core and broad shallow wings, which affect the intrinsic profiles in a non-trivial way.

To get the final predicted absorption-line profile, one needs to convolve the predicted absorption profile given by Eq.~1 with the LSF, i.e.\
\begin{equation}
f_{\rm abs}^{\rm obs}(v)= f_{\rm abs}(v)\ast {\rm LSF}(v). \label{Iapp}
\end{equation}
In Figure 4 we show simulations of $f_{\rm abs}^{\rm obs}(v)$ (right panels) for a range of input $f_{\rm abs}(v)$ (left panels). All simulations are with a $b$-parameter of 50~\kms. In the upper panels we explore the effect of \nion, ranging from optically thin ($N_{\rm ion}=10^{13}$~\cm) to highly saturated ($N_{\rm ion}=10^{20}$~\cm) lines, with CF=1 in all cases. In the lower panels we explore the effect of the CF, for CF=0.1 to 1, for mildly saturated ($N_{\rm ion}=10^{18}$~\cm) and highly saturated ($N_{\rm ion}=10^{20}$~\cm) lines. The atomic parameters used for the synthetic absorption line are $\lambda_{\rm ion}=1200$~\AA, $\Gamma=2.5\times10^8$~s$^{-1}$, and $f_{\rm os}=1$, where $\Gamma$ is the spontaneous radiative transition probability, and $f_{\rm os}$ is the oscillator strength. The chosen values are typical for the UV absorption lines analyzed in this paper. The STIS LSFs are plotted in panel (b), as tabulated at the NASA/STScI \hst\ STIS database\footnote{\url{http://www.stsci.edu/hst/stis/performance/spectral\_resolution/}}.
\placefigure{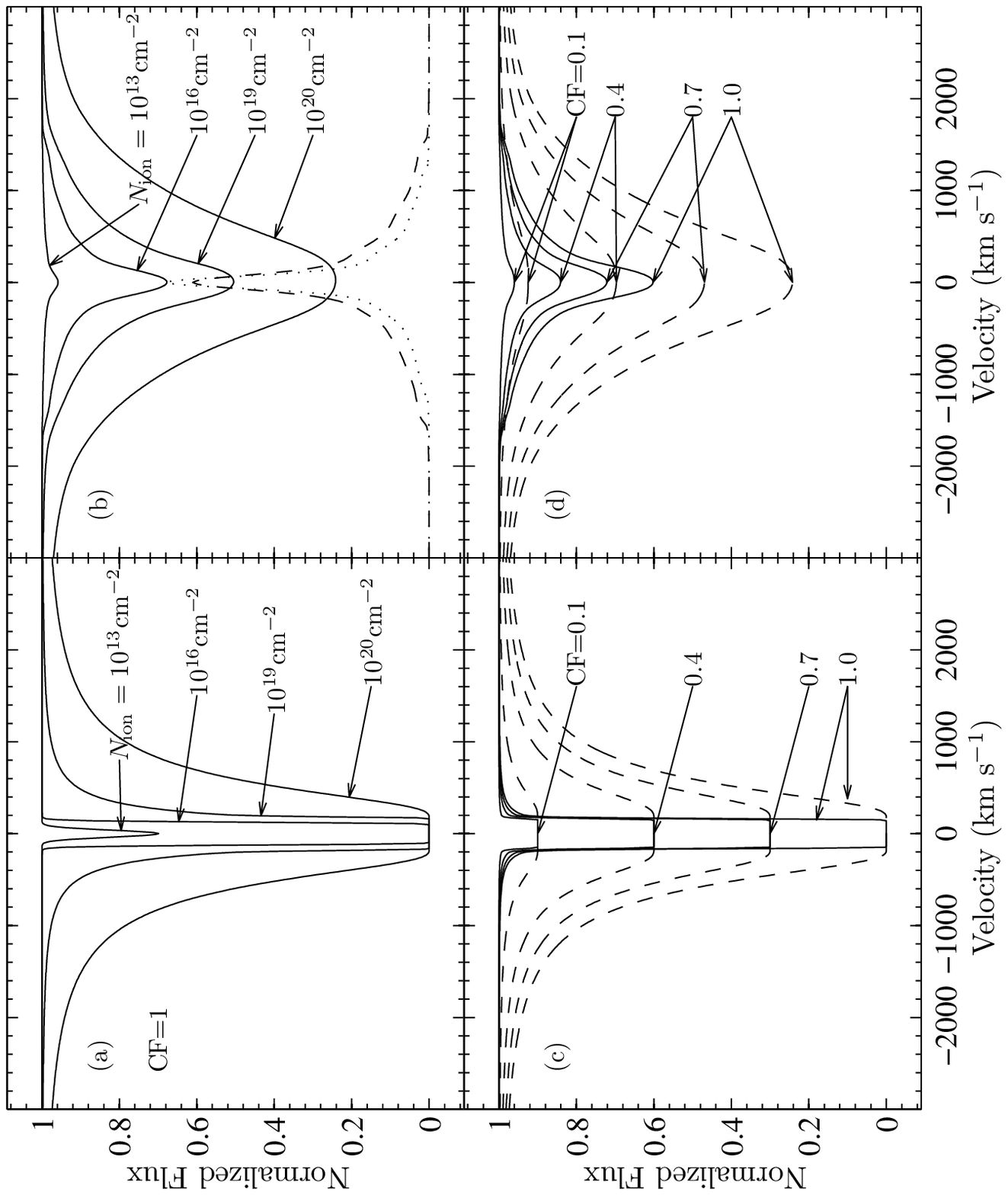}

The upper panels of Fig.~4 demonstrate that saturated lines will generally appear unsaturated, even when $f_{\rm abs}(v)$ is significantly broader than LSF$(v)$. For example, the $N_{\rm ion}=10^{19}$~\cm\ case, where $\tau\gg 1$ within 200~\kms\ of the line center in $f_{\rm abs}(v)$ [panel (a)], appears as a $\tau<1$, FWHM=1000~\kms, absorption profile in $f_{\rm abs}^{\rm obs}(v)$ [panel (b)]. This strong distortion results from the extended broad wings of the LSF, which bring in unabsorbed continuum flux to the highly saturated line core. Even the highly saturated absorption at $N_{\rm ion}=10^{20}$~\cm\ [panel (a)], where the intrinsic FWHM=1000~\kms\ (five times the LSF), appears unsaturated following a convolution with the LSF [panel (b)]. The lower panels of Fig.~4 show cases where the intrinsic profiles appear unsaturated due to partial coverage. Comparison of panels (b) and (d) shows that it is practically impossible to differentiate a range of \nion\ at a given CF, from a range of CF at a given \nion.

Below we therefore adopt two extreme approaches for modeling the absorption-line profiles:
\begin{enumerate}
\item Assume $b_{\rm bulk}=0, b_{\rm thermal}=10$~\kms, which corresponds to cold gas with $T\sim10^4$~K producing the minimal possible velocity dispersion. This assumption leads to the highest possible $N_{\rm ion}$, where the absorption is dominated by the Lorentzian wings. We cannot measure the value of $b$ because it is well below the spectral resolution.
\item Assume absorbing gas that has the maximum possible $b_{\rm bulk}$ consistent with the observed absorption-line profiles. This assumption allows an optically thin absorber, and leads to the lowest possible $N_{\rm ion}$.
\end{enumerate}
Since the observed absorption-line profiles are dominated by the instrumental LSF, we cannot resolve any velocity dependence of CF$(v)$, and we therefore use a velocity-independent CF while reconstructing the spectra. Although much higher resolution spectra ($\sim10$~\kms) obtained with the STIS echelle, are presented by Crenshaw et~al.\ (2004), the low S/N of these spectra does not allow significant constraints on the absorption width.

\subsection{The reconstruction technique}
The common technique for measuring $f_{\rm abs}^{\rm obs}(v)$ is by interpolating the unabsorbed spectrum between the two nearest unabsorbed points, where $f_{\rm abs}^{\rm obs}(v)$ is then calculated by dividing the measured absorbed spectrum by the fitted unabsorbed spectrum. This method may be rather uncertain when the absorption occurs in regions affected by emission lines, as the unabsorbed spectrum is difficult to estimate due to the possible presence of emission features whose shape is not well constrained. To avoid the possible significant systematic errors in  $f_{\rm abs}^{\rm obs}(v)$ we adopt a reverse approach of spectral ``reconstruction''. We calculate $f_{\rm abs}^{\rm obs}(v)$ based on a certain model for the absorber, divide the observed spectrum by $f_{\rm abs}^{\rm obs}(v)$, and get the reconstructed unabsorbed spectrum. The parameter models are deemed acceptable if the reconstructed spectrum appears ``acceptable''. This method is formally not more accurate than the standard approach, where $f_{\rm abs}^{\rm obs}(v)$ is directly measured, however it should allow a larger range of reconstructed spectral shapes, and thus a more realistic estimate of the possible uncertainty in the fit parameters. It should be noted that the adopted approach also allows to estimate the true range of \nion, without introducing a theoretical bias, because $f_{\rm abs}^{\rm obs}(v)$ is produced using atomic physics only, without assuming a specific photoionization model or particular elemental abundances.

Since the absorption-line features appear to be generally unresolved, we attempt only to constrain the range of possible values for \nion\ and CF. We assume a set of CF and $b$, and then reconstruct the observed absorption-line profiles by varying \nion\ only (the algorithms are described below). The assumed sets of CF and $b$ are:
\begin{description}
  \item[{\boldmath${\rm CF}=1$}] Constrain the column-density for a full coverage absorber.
  \begin{itemize}
    \item Measure the lower-limit of \nion\ assuming $b>100$~\kms\ -- we iterate over different values of $b$, while reconstructing simultaneously the most prominent absorption-lines in our spectra: \ion{Si}{2}~$\lambda1260.42$, \ion{Si}{2}~$\lambda1526.71$ and \ion{C}{2}~$\lambda1334.53$, using the same column-density for both \ion{Si}{2} lines. A $b$-parameter of 180~\kms\ produces a featureless corrected spectrum for the above absorption lines for both the low- and high-states. A larger value of $b$ produces absorption-line profiles which are broader than those observed. This value of $b$ is also adopted for the weaker absorption features and for the high-ionization species absorber.
    \item Measure the upper-limit of \nion\ using  $b=10$~\kms.
  \end{itemize}
  \item[{\boldmath${\rm CF}<1$}] Constrain the column-density for the lowest possible CF.
  \begin{itemize}
    \item Measure the lower-limit of \nion\ using $b>100$~\kms\ -- as above, $b=180$~\kms\ produces featureless corrected spectra for the prominent absorption lines.
    \item Measure the upper-limit of \nion\ using $b=10$~\kms\ -- the lowest best fit for all ions and all possible locations of the absorbing gas is ${\rm CF}\simeq1$. Forcing ${\rm CF}<1$ for a $\tau\gg1$ column leads to a box-shaped absorption profile, which cannot reproduce the observed absorption, unless we allow ${\rm CF}\simeq1$. Thus, the results are identical to the above case for ${\rm CF}=1$.
  \end{itemize}
\end{description}

We identified three distinct absorption systems \Al, \Ah, and B (see Figs.\ 2 and 3), having $v_{\rm shift}$ of $-250$, $-250$, and $-840$, respectively. While systems \Al\ and \Ah\ have prominent components in most of the low- and high-ionization absorption lines, respectively, system B is clearly detected only in \ion{C}{4} and probably in \ion{N}{5}, both in the high-state. The value of $v_{\rm shift}$ of systems \Al\ and \Ah\ is found using the prominent absorption lines, and the value of $v_{\rm shift}$ of system B is found using \ion{C}{4} doublet at the high-state. It should be noted that the distinction between systems \Al\ and \Ah\ is physically motivated and is not observationally required. The reconstruction procedure also depends on the assumed physical location of the absorbing gas, as further described below.

\subsection{The physical location of the absorbing gas}
There are three possible distinct locations for the absorbing gas. (i) Outside the NLR, in which case it is a pure foreground absorber which absorbers all the observed emission. (ii) Between the NLR and the BLR, in which case the NLR is not absorbed. (iii) Between the BLR and the continuum source, in which case only the continuum source is absorbed. In cases (ii) and (iii) variations in the emission of the continuum and the BLR, relative to the emission of the NLR, can lead to changes in the observed absorption profiles, even when the absorber is non-variable. Below we describe the reconstruction procedure for the different possible absorber locations.

\subsubsection{Absorber located outside the NLR}
The reconstruction procedure for a foreground absorber is as follows:
\begin{enumerate}
   \item Choose a specific set of $b$ and CF.
   \item Assume \nion\ and use it for all absorption lines of that ion for a given absorption system.
   \item Calculate $\tau$ using $\tau(v)=N_{\rm ion}\frac{\sqrt{\pi}e^2}{m_{\rm e}c}f_{\rm os}\frac{\Psi(v)}{\Delta\nu_D}$, where $e$ and $m_{\rm e}$ are the charge and mass of electron, $c$ is the speed of light, and $\Delta\nu_D=\nu_{\rm ion}\times b/c$ is the line width ($\nu_{\rm ion}=c/\lambda_{\rm ion}$). The calculation is made using a velocity scale, and is transformed to a wavelength scale.
   \item Transfer the wavelength scale of $\tau(\lambda)$ to the rest frame of the relevant component by adding $\lambda_{\rm shift}=v_{\rm shift}/c\times\lambda_{\rm ion}$ to $\lambda$ i.e., $\lambda\rightarrow\lambda+\lambda_{\rm shift}$.
   \item Calculate the EW of the absorption-line using $\tau(\lambda)$ from step (3), and CF from step (1).
   \item If there is another absorption line or absorption system within a range of $\sim20$~\AA\  (approximately twice the maximal LSF extended wings), repeat steps (1)--(6). Calculate $\tau_{\rm tot}(\lambda)$, the total optical depth, by summing over all individual $\tau(\lambda)$ from step (3), $\tau_{\rm tot}(\lambda)=\sum_k\tau_{{\rm ion}(k)}(\lambda)$.
   \item Calculate $f_{\rm abs}(\lambda)$ using $\tau_{\rm tot}(\lambda)$ from step (6), and CF from step (1). Note that CF is the same for all absorption features.
   \item Calculate $f_{\rm abs}^{\rm obs}(\lambda)$ by convolving $f_{\rm abs}(\lambda)$ from step (7) with the LSF. For $\lambda\leq1350$~\AA\ we use the LSF provided at 1200~\AA, and for $\lambda>1350$~\AA\ we use LSF provided at 1500~\AA.
   \item Iterate over steps (1)--(8) until the corrected spectrum is either featureless or has plausible emission feature (e.g., narrow-line emission).
\end{enumerate}
The reconstructed spectra and the modeled absorption profiles are presented in Figure 5. The measured \nion\ and EW are tabulated in Table 1. It should be noted that although the \ion{C}{4} absorption feature attributed to system \Ah\ is only loosely constrained\footnote{A relatively large amount of flux can be added to the \ion{C}{4} feature throughout the reconstruction process, while keeping the feature nearly symmetrical (a small blue-shift is introduced).} for the high-state, it is very unlikely that the real \nion\ values are an order of magnitude larger than those tabulated. Such columns produce very strong \ion{C}{4} in the reconstructed spectrum, which is not detected in other emission-lines  [e.g., \ion{Si}{4}; see Fig.\ 1, panel (a1)]. \placefigure{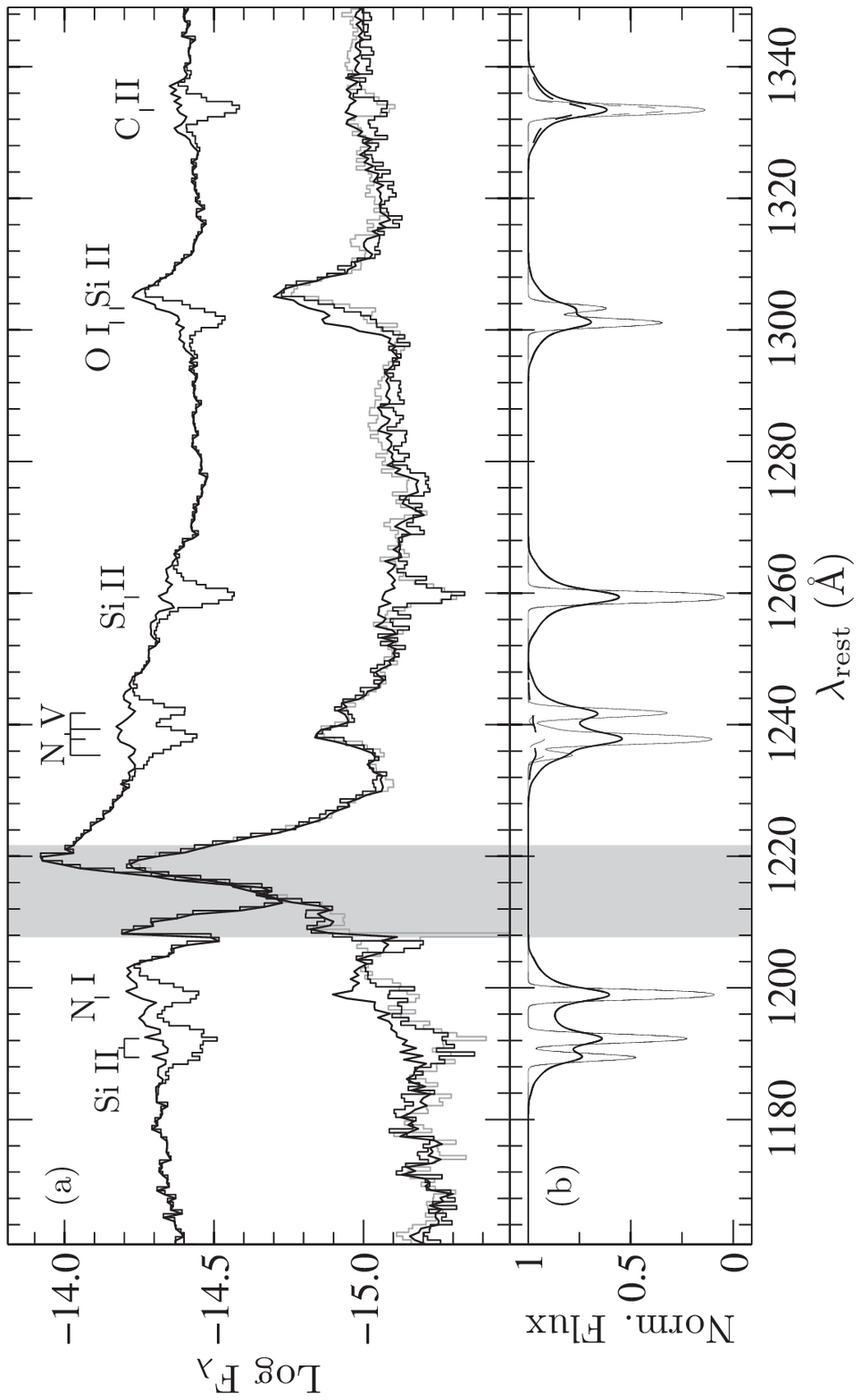,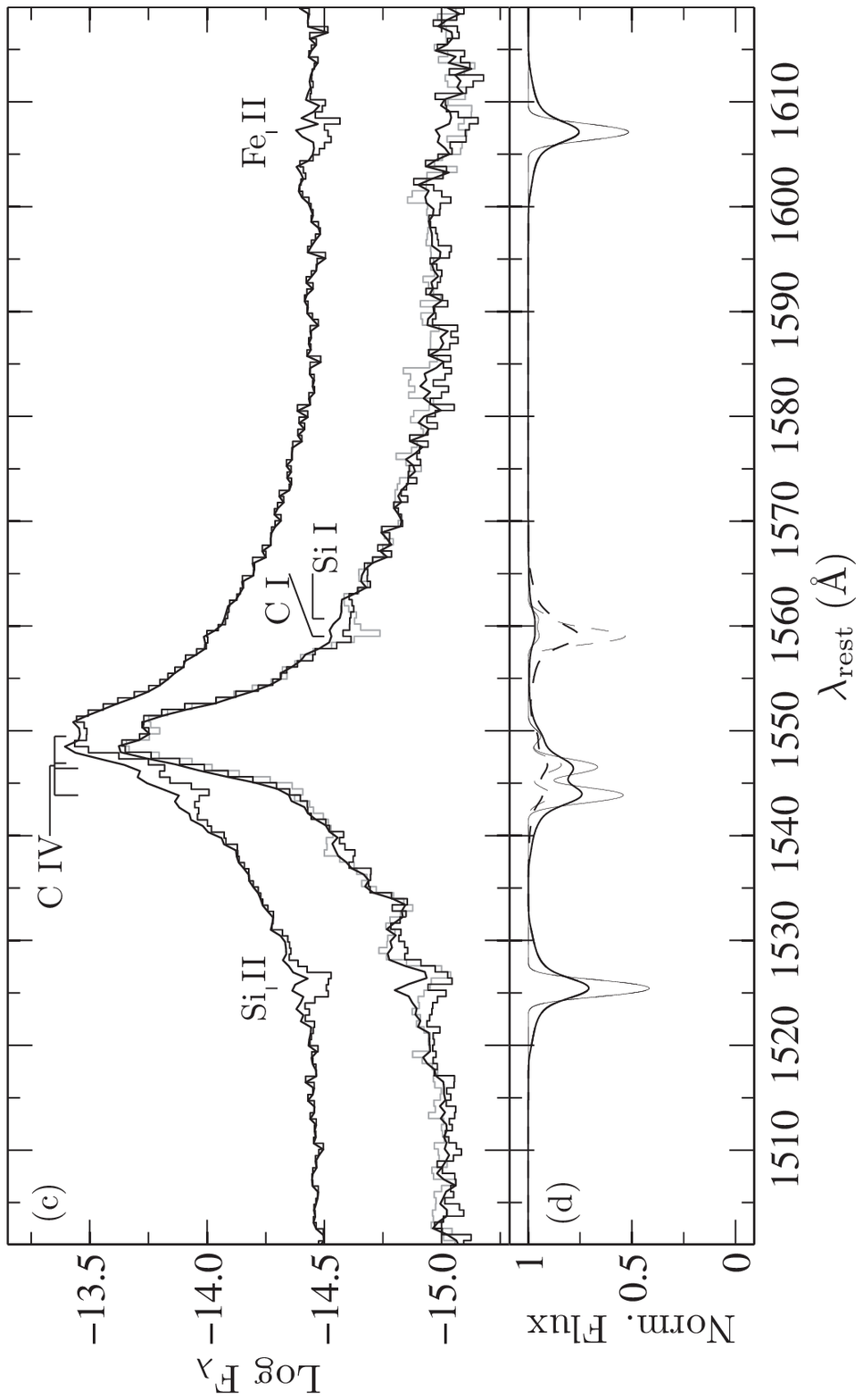}

\subsubsection{Absorber located between the NLR and BLR or inside the BLR}
Since the NLR is not absorbed, we need to estimate the contribution of the NLR to the observed emission lines. Crenshaw et~al.\ (2004), clearly detected narrow \ion{C}{4} emission having FWHM of $\sim100$~\kms\ (see Fig.\ 2, there). These narrow components are not resolved in the lower resolution STIS spectra, and we therefore have to get an indirect estimate of the NLR emission component. A FWHM of $\sim100$~\kms\ indicates that the lines are produced at a significantly larger distance from the center, compared with the BLR, and would therefore vary on a significantly longer timescale. Below we derive the NLR emission spectrum based on the assumption that it did not vary between the low- and high-states. If the absorber is located inside the BLR, then only the continuum is absorbed. Since the continuum is a featureless power-law, it is simple to estimate its shape at any state. The specific procedure for the spectral reconstruction is detailed below.
\begin{enumerate}
   \item Obtain a continuum plus BLR emission spectrum. This is done by subtracting the low-state spectrum from the high-state spectrum [e.g.\ Fig.\ 1, panel(d)], thus subtracting out the NLR line-emission, which we assume does not vary between the low- and high-states.
   \item Obtain the net continuum emission. This is done by fitting a local power-law continuum, $F_{\rm cont}$, between 1150~\AA\ and 1280~\AA\ for the \ion{N}{5} doublet and 1500~\AA\ and 1595~\AA\ for the \ion{C}{4} doublet, where the continuum level is set by the averaged flux in region within $\pm5$~\AA\ range of the bounding wavelengths.
   \item Correct for absorption. We use a procedure similar to that described in \S~3.4.1. For absorption of the continuum only, correct the observed flux, $F_{\rm obs}(\lambda)$, using the calculated $f_{\rm abs}^{\rm obs}(\lambda)$ [see step (8) in \S~3.4.1] in the following manner: $F^{\rm corr}_{\rm obs}(\lambda)= F_{\rm obs}(\lambda)+(1-f_{\rm abs}^{\rm obs})\cdot F_{\rm cont}$. The same parameters are used for both absorption systems \Ah\ and B.\footnote{A somewhat better fit of \ion{N}{5} can be reached using $v_{\rm shift}=-230$~\kms\ instead of $-250$~\kms\ for system \Ah, but both values are consistent within the uncertainties.} The lowest possible CF for system \Ah\ is constrained using the \ion{N}{5} feature. The \ion{C}{4} feature is used to constrain the CF for system B. When different CFs are adopted for the two absorption systems, a multiplication of the two individual intrinsic absorption profiles, associated with the two systems, is used to calculate the total intrinsic absorption profile i.e., $f_{\rm abs}^{\rm tot}(\lambda)=\left\{\left(1-{\rm CF_{\rm A}}\right)+{\rm CF_{\rm A}}\cdot\exp{\left[-\tau_{\rm tot}^{\rm A}(\lambda)\right]}\right\}\times \left\{\left(1-{\rm CF_{\rm B}}\right)+{\rm CF_{\rm B}}\cdot\exp{\left[-\tau_{\rm tot}^{\rm B}(\lambda)\right]}\right\}$, where ${\rm CF_{\rm A}}$ and ${\rm CF_{\rm B}}$ are the adopted CFs for systems \Ah\ and B, respectively.
   \item Obtain the NLR emission. We produce a synthetic narrow \ion{C}{4} and \ion{N}{5} doublets based on the \ion{He}{2} narrow line. First, estimate the width of the the narrow-line emission by measuring the \ion{He}{2}~$\lambda1640.46$ profile in the low-state. The measured intrinsic FWHM of \ion{He}{2} is 400~\kms, and it is possibly unresolved. Allow a wavelength-shift of the synthetic doublet relative to the rest-frame of the object (the doublet is shifted by $-140$ and 0~\kms\ for \ion{N}{5} and \ion{C}{4}, respectively, in the current analysis). The flux of the doublet is a free parameter, while the ratio between the two lines is governed by the ratio of their $g\cdot f_{\rm os}$ parameters (adopted from Morton 1991), where $g$ is the statistical weight of the lower level.
   \item Vary the absolute flux of the \ion{N}{5} or \ion{C}{4} He-like doublet and subtract it from the low spectrum. Correct the resulting spectrum for absorption using the absorption profile measured in step (3). When an absorption of continuum only is assumed, first fit a local power-law continuum as prescribed in step (2). Iterate over this step until the reconstructed spectrum is featureless or has an expected emission feature.
\end{enumerate}
The values of \nion\ and CF for ${\rm CF}<1$ are measured by repeating steps (1)--(3) and adopting the narrow-emission doublets found for ${\rm CF}=1$. The \nion\ values measured for ${\rm CF}<1$ (see Tables 2 and 3) are lower-limits. They can be as much as $\sim3$ times higher, and still produce a reasonable absorption-line profile reconstruction. The lowest possible CF assuming that the absorber is located outside the BLR are 0.7 and 0.5 for systems \Ah\ and B, respectively. The lowest possible CF assuming that the absorber is located between the BLR and the continuum source are 0.8 and 0.5 for systems \Ah\ and B, respectively. Those values are constrained using \ion{N}{5} only, because the \ion{C}{4} absorption feature cannot be reconstructed while assuming absorption of the continuum only (see below). The \ion{C}{4} absorption feature attributed to system \Ah\ is loosely constrained for all values of CF, as found above for the case of a foreground absorbing screen. The strong \ion{C}{4} narrow-line emission contributes to the difficulty of constraining the \ion{C}{4} absorption-line profile measurements.

Figures 6 and 7 present the reconstructed \ion{N}{5} and \ion{C}{4} features in the high$-$low state, which represents the net continuum plus BLR emission, and in the low spectra, assuming a fixed absorption-line profile for both states. The unabsorbed narrow-line emission has a total EW of 8.5 and 75~\AA\ for the \ion{N}{5} and \ion{C}{4} doublets, respectively. The left panels in Fig.~6 present the case where both the continuum and the broad-line emission are corrected for absorption, and the right panels present the case where only the continuum is corrected for absorption. The \ion{C}{4} features cannot be reconstructed assuming that only the continuum is absorbed, as the continuum is so weak that even a complete absorption of it cannot explain the observed depth of the absorption features. Even when the synthetic profile is saturated, only a fraction of the observed \ion{C}{4} absorption is reconstructed, due to the small contribution of the continuum to the total emission in this wavelengths range (Fig.\ 7). Thus, we present in Figure 7 only the reconstructed \ion{C}{4} feature assuming an absorber located outside the BLR. As we show below, a physically motivated model indicates that the absorption profile may differ somewhat in the low- and high-states, and the procedure for this case is described in \S~4.
\placefigure{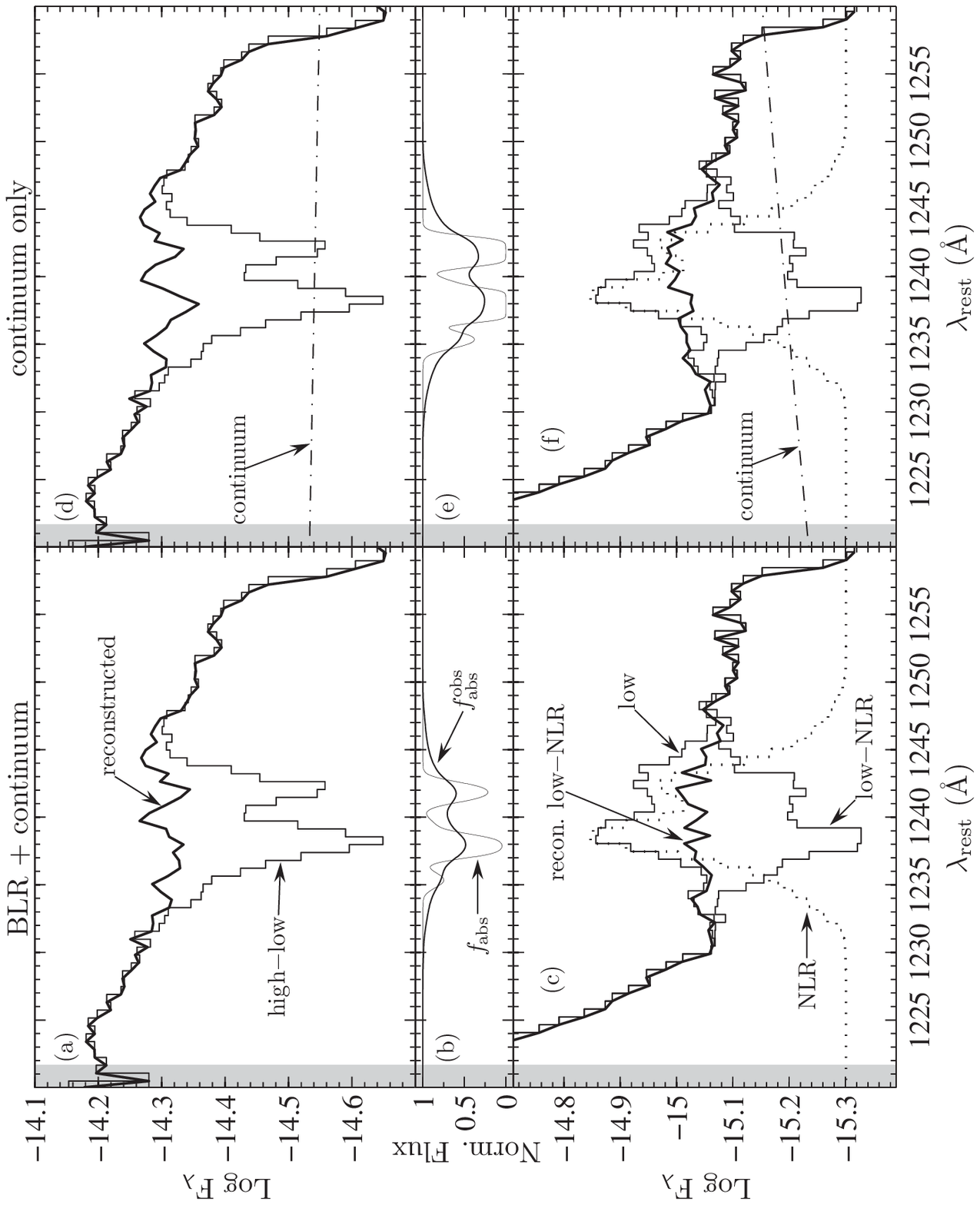}\placefigure{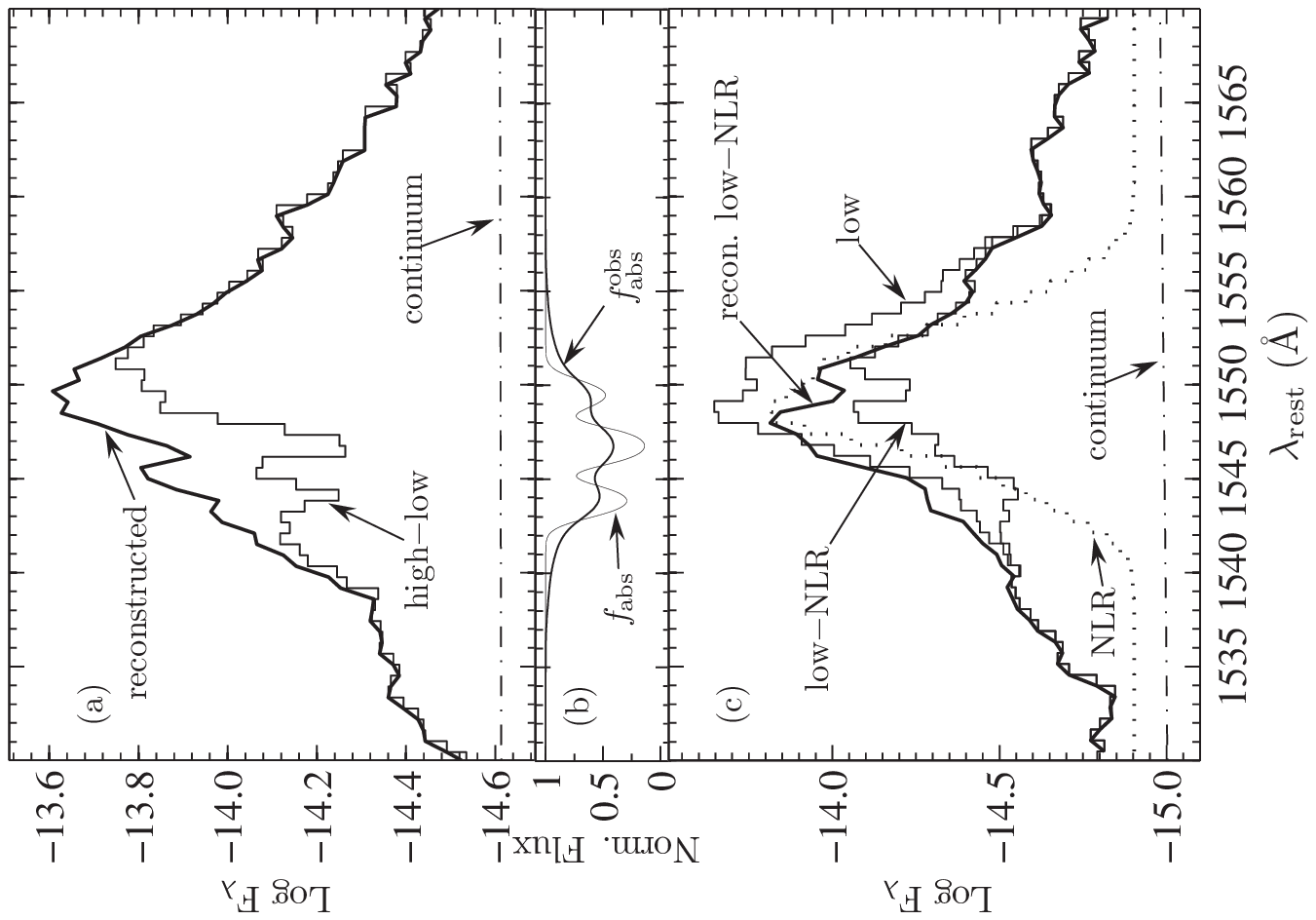}

\subsubsection{Results}
Table 1 presents the measured \nion\ and EW for systems \Al, \Ah, and B during the low- and high-states for different sets of CF and $b$, for an absorber outside the NLR. Column 1 lists the ion. Column 2 lists the laboratory wavelength of the absorption line. Column 3 lists the flux state, high or low; or both if no state is listed. Columns 4, 6, and 8 list \nion\ for the following three sets of CF and $b$ values; ${\rm CF}=1$, $b=180$~\kms; ${\rm CF}=1$, $b=10$~\kms; and ${\rm CF}=0.75$, $b=180$~\kms. The first two sets correspond to the lower and upper limits on the columns, and the third set demonstrates the effect of a reduced CF. Columns 5, 7, and 9 list the absorption EW for these three sets (the possible presence of an additional absorption system is discussed in \S~5.1.4). Tables 2 and 3 are identical to Table 1, but with either an unabsorbed NLR (Table 2), or continuum absorption only (Table 3). Note that the absorption EW refers to the absorbed component only, and not to the absorption EW of the observed feature, which is diluted by the unabsorbed NLR emission (Table 2), or unabsorbed NLR plus BLR (Table 3). Although there may be a weak absorption in the $\sim1400$~\AA\ blend of \ion{Si}{4} and \ion{O}{4}] emission (Fig.~3), we did not try to measure it, as some of the apparent absorption features could be an artifact of the complex emission nature of this blend. We have tried to reconstruct the blend with a synthetic He-like profile \ion{Si}{4} doublet and \ion{O}{4}] quintet, using the measured FWHM of \ion{He}{2}, but the reconstruction failed, implying that the profiles of the individual features are probably not He-like.
\placetable{measN}\placetable{measNBLR}\placetable{measNCont}

\section{MODELING OF THE ABSORBING GAS}
We use the photoionization code {\sc cloudy} v.\ 06.02b (Ferland et~al.\ 1998) to derive the physical properties of the absorbers, i.e.\ Hydrogen column-density $\Sigma$, the electron density $n_{\rm e}$, and the ionizing flux (through the ionization parameter $U$), or equivalently the distance $R$ of the absorber from the ionizing source, which reproduce \nion\ measured above from the spectral reconstruction. The assumed spectral energy distribution (SED) for the high- and low-states, are presented in Tables 4 and 5. Note that the SED during the low-state is harder than during the high-state. The mean ionizing photon energy $\langle h\nu \rangle$ is 6.8~Ry for the high-state SED, and 35.3~Ry for the low-state SED, where the spectrum is very hard. Finally, we assume solar abundance throughout the analysis, as expected for objects found at the low-end of $M_{\bullet}$ (Hamann et~al.\ 2007, and references therein), although Kraemer et~al.\ (1999) deduce sub-solar abundances for \ngc. The exclusion of the \Lya\ wavelength-region from the analysis (see \S~2.2), prevents a direct constraint on the gas metallicity using the Hydrogen to metals column ratio.
\placetable{SEDhigh}\placetable{SEDlow}

The calculations are made for a grid of models with $-4\leq\log U\leq 1$ for the low-state and $-3\leq\log U\leq 1$ for the high-state (in steps of 0.5 in $\log U$), $5\leq\log n_{\rm e}\leq 13$ (in steps of 1), and $18\leq\log\Sigma\leq 24$ (in steps of 1). The grid is then interpolated for each ionic column using a spline approximation in steps of 0.2, 0.5, and 0.5~dex for $U$, $n_{\rm e}$, and $\Sigma$, respectively. It should be noted that \nion\ is mostly set by $U$ and $\Sigma$, and is only weakly dependent on $n_{\rm e}$. The absorber parameters are determined by the following procedure.

\begin{enumerate}
   \item Find the possible range of EW from the spectral reconstruction technique, for a specific absorption line for a given absorber location and CF, as described in \S~3. We adopt an uncertainty of 10\% on the measured EW range.
   \item Assume $b$ and calculate the curve of growth for the given absorption line.
   \item Find the range of \nion\ which reproduces the EW range found in step (1) using the curve of growth from step (2). If only an upper limit on the EW is available, find the corresponding upper limit on \nion.
   \item If more than one absorption line is attributed to a given ion, repeat steps (1)--(3) for all lines. Average the independent determinations of the upper and lower limits on \nion.
   \item Repeat steps (2)--(4) for different values of $b$. We iterate from 10 to 200~\kms\ in steps of 0.1~dex.
   \item Repeat steps (1)--(5) for all of the potentially significant expected absorption lines, and the observed absorption lines from a particular absorption system.
\end{enumerate}
Finally, the sets of parameters ($U$ and $\Sigma$) and the range of $n_{\rm e}$ which reproduce simultaneously \nion\ of the different ions are estimated for each value of $b$ for the high- and low-states. This procedure is not performed for the low-ionization species absorber, because as we show below (\S~5.1.1) this absorber is probably not intrinsic to the \ngc\ nucleus.

The distance of the absorber from the ionizing source in the low- and high-states is determined as follows. By definition $U\equiv n_\gamma/n_{\rm e}$, where the ionizing photon density is given by $n_\gamma=L_{\rm ion}/4\pi R^2 c \langle h\nu \rangle$, where $L_{\rm ion}$ is the ionizing luminosity and $R$ is the distance from the ionizing source; thus, $R=L_{\rm ion}^{1/2}(4\pi c\langle h\nu \rangle U n_{\rm e})^{-1/2}$. As described above, $\langle h\nu \rangle$ equals 35.3 and 6.8~Ry for the low- and high-state, respectively. Using {\sc cloudy} we also find the relation between the ionizing luminosity and the UV luminosity at 1350~\AA, $L_{\rm ion}/\nu L_{\nu}(1350\mbox{~\AA})\approx$ 19.8 and 5.5 for the low- and high-state SED, respectively. Finally, we adopt a distance $D=4.3$~Mpc (Thim et al.\ 2004), which gives $\nu L_{\nu}(1350\mbox{~\AA})\approx$ 2.5 and $9.3\times10^{38}$~ergs~s$^{-1}$ for the low- and high-states. We transfer the sets of $U$, $\Sigma$ and $n_{\rm e}$ found above for the high- and low-states into sets of $R$, $\Sigma$ and $n_{\rm e}$, as all these quantities describe the absorber independently of the ionizing continuum.

We then search for matching sets of $R$, $n_{\rm e}$, and $\Sigma$ in the high- and low-states solutions, with $R$ values which agree within 30\%, for either a foreground absorber or an absorber inside the NLR. If such set of parameters is found, we deduce that there might be a non-variable absorption system located at a distance $R$, having the estimated $n_{\rm e}$ and $\Sigma$. If no set is found assuming a foreground absorber, then this assumed location for a non-variable absorption system can be ruled out.

The assumption of a constant absorption line profile in the low- and high-states is oversimplified, as the large change in ionizing flux is likely to change the ionization state of the absorber. In order to remove the phenomenological constraint of a constant absorption-line profile, the following procedure is used. We adopt \nion\ measured for the high$-$low spectrum using the procedure described in \S~3.4.2. We then estimate the $R$, $n_{\rm e}$, and $\Sigma$ as described above, while using the SED of the high-state. These parameters are assumed to hold in the low-state, and are used to calculate $U$ for this state, which is then used together with $n_{\rm e}$ and $\Sigma$ to estimate \nion\ in this state. Finally, we check whether the estimated \nion\ can successfully reconstruct the low-state spectrum, while altering the narrow-line emission flux. The success of the reconstruction is determined by eye. It should be noted that we do not try to refine the measured \nion\ and the estimated physical parameters, including the $b$-parameter, by searching for the ``best'' reconstruction of the low-state spectrum, due to the rather limited S/N and the broad LSF of the instrument. We rather just check the consistency of the adopted solutions. Sets of $R$, $n_{\rm e}$, and $\Sigma$ which produce a successful reconstruction, assuming either variable or non-variable absorption-line profile, indicate a constant absorber inside the NLR. As described below, this is the most plausible location for the observed high-ionization absorbers (systems \Ah\ and B).

\section{ORIGIN OF THE ABSORPTION SYSTEMS}
The main question that should be answered is whether the absorption originates close to the nucleus of \ngc\ or whether it is an unrelated foreground system. Since $n_{\rm e}$ in absorption studies is generally poorly constrained, the distance may be uncertain by a few orders of magnitude (e.g.\ Crenshaw et~al.\ 2003). However, here we use the absence of absorption from metastable levels of \ion{Si}{2} to show that system \Al\ must occur outside of the NLR of \ngc, most likely in the ISM of the host galaxy or possibly in the Milky Way. Further, based on the large amplitude of the continuum variability, and the observed change in the \ion{N}{5} and \ion{C}{4} absorption profiles, we show that the high-ionization absorber (systems \Ah\ and B) must occur inside the NLR, and most likely outside the BLR of \ngc. In \S\S~5.1 and 5.2 below we describe the constraints as a function of the assumed location, for each of the absorption systems, and in \S~5.3 we compare our results to earlier studies.

\subsection{Absorbers outside the NLR}
\subsubsection{System \Al\ -- nuclear or foreground?}
System \Al\ ($v_{\rm shift}=-250$~\kms) of the low-ionization lines, is seen prominently in \ion{C}{2}, \ion{N}{1}, \ion{O}{1}, \ion{Si}{2} and \ion{Fe}{2} ions during both flux-states and in \ion{C}{1} and \ion{Si}{1} ions during the low-state. Most of the absorption-lines do not vary between the low- and high-sates. The exceptions are the \ion{C}{1}~$\lambda1560.31$ + \ion{Si}{1}~$\lambda1562.00$ absorption blend and \ion{C}{2}~$\lambda1334.53$ absorption line, which are the only two low-ionization absorption features that appear to vary between the low- and high-states. Although the \ion{C}{1}~$\lambda1560.31$ + \ion{Si}{1}~$\lambda1562.00$ blend appears significant, its width in the low 2 state is below the spectral resolution (Fig.~2), and it is thus most likely an artifact. The \ion{C}{2}~$\lambda1334.53$ absorption in the low 2 state is consistent with the absorption in the high-state, and we suspect that the weaker absorption in the low 1 state may be affected by artifacts as well. The absorption profiles of the prominent low-ionization species, \ion{Si}{2}, \ion{N}{1}, and \ion{O}{1}, can be reconstructed using the same parameters for both the low- and high-state spectra, consistent with no variability of the low-ionization species absorber.

The basic question is whether the low-ionization system \Al\ lines are intrinsic to the \ngc\ nucleus or, whether they can be attributed to the Galactic ISM or the ISM of \ngc. The velocity-shift of $-250$~\kms\ is quite close to $v_{\rm shift}\simeq-320$~\kms, the redshift of \ngc. We first check whether \nion\ measured here are consistent with the expected Galactic ISM values. Murphy et~al.\ (1996) report a column-density of $N_{\rm H}=14.3\times10^{19}$~\cm\ towards \ngc\ based on measurements of the 21~cm \ion{H}{1} line. Using the solar abundances from Morton (1991, table 1) the expected upper-limits on the \ion{C}{2} and \ion{Si}{2} \nion\ are $5.2\times10^{16}$ and $5.1\times10^{15}$~\cm, respectively. The \nion\ of \ion{C}{2} and \ion{Si}{2} measured for system \Al\ assuming optically-thin gas ($b=180$~\kms), are below those upper-limits. We note in passing that Savage et~al.\ (2000) do not detect Galactic \ion{C}{1} and \ion{Si}{1} absorption-lines in their \hst\ quasar absorption line key project. This points against an ISM origin for these lines, and suggests they are either formed in a different region, or are possibly an instrumental artifact.

The strongest constraint on the location of system \Al\ comes from the absence of absorption lines attributed to metastable levels, in particular the \ion{Si}{2}$^\ast$ $\lambda\lambda1264.74, 1265.02$ doublet. These absorption lines are produced by an electron in the excited $^3P_{3/2}$ level, 287.24~cm$^{-1}$ above the ground level, to an upper level which is split into two sub-levels, $^3D_{5/2}$ and $^3D_{3/2}$. Adopting $\Gamma=2.17\times10^{-4}$~sec$^{-1}$ and a collision strength of 5.7 for the $^3P_{3/2}$ level (Pradhan \& Peng 1995)\footnote{For a temperature of $10^4$~K, a typical value for photoionized gas.}, we get a critical density of $\sim1.8\times10^3$~cm$^{-3}$. Thus, the absence of the \ion{Si}{2}$^\ast$ absorption lines implies that the density of the absorbing gas is significantly lower than $\sim1.8\times10^3$~cm$^{-3}$. The likely value of $U$ is below $10^{-3}$, otherwise Si is ionized beyond \ion{Si}{2}. These limits on $n_{\rm e}$ and $U$ place the absorber at a distance larger than the ``OUTER'' NLR component, identified by Kraemer et~al.\ (1999) in \ngc. A plausible location for the \Al\ absorber is the ISM of of \ngc\ or that of the Galaxy. Crenshaw et~al.\ (2004) resolve two low-ionization absorption components, one of which they attribute to the Galactic ISM and the other to the ISM of the host. The velocity-shift difference between those two components is $\sim300$~\kms\ which is below the current spectral resolution. Thus, system \Al\ may be a blend of absorption arising in two distinct absorption systems found in the ISM of the Galaxy and of the host.

\subsubsection{Are systems \Al\ and \Ah\ connected?}
System \Ah\ ($v_{\rm shift}=-250$~\kms) is seen prominently in the \ion{N}{5} doublet during the high-state, and is not seen in the \ion{Si}{4} ions during both flux states. The system \Ah\ absorption seen in \ion{N}{5} at the high-state must be attributed to a distinct absorption system (i.e., other than system \Al) for the following reason. Unlike the low-ionization lines, the \ion{N}{5} absorption is not detected in the low-state, ruling out a foreground screen with a fixed absorption profile. Could the absorber have fixed properties, but variable ionization, such that N is ionized to \ion{N}{5} in the three month between the low- to high-state, explaining the appearance of the \ion{N}{5} absorption in the high-state? The H ionizing photon flux in the location where $U<10^{-3}$ and $n_{\rm e}<10^3$~cm$^{-3}$ is $cn_{\rm e}U<3\times 10^{10}$~s$^{-1}$~cm$^{-2}$. Only photons above 5.7 Ry can photoionize \ion{N}{4} to \ion{N}{5}, reducing the above flux by a factor of about 5.7. The absorption cross section of \ion{N}{4} at the ionization threshold is $10^{-18}$~cm$^{-2}$ (e.g.\ Osterbrock 1989), leading to an ionization timescale of $>(5\times10^9\cdot 10^{-18})^{-1}$~s, i.e.\ $>6$ years. Thus, \ion{N}{4} could not have been photoionized to \ion{N}{5} in three months at the location of the low-ionization absorber. As we show below, the \ion{N}{5} system \Ah\ absorption must arise between the BLR and NLR of \ngc.

\subsubsection{System B}
System B ($v_{\rm shift}=-840$~\kms) is seen prominently in the \ion{C}{4} doublet during the high-state. It also appears to be present in the \ion{N}{5} doublet during the high-state, but it is blended with the stronger system \Ah\ doublet absorption. It is not seen in lower-ionization ions, including \ion{Si}{4}. As argued above (\S~5.1.2) for the system \Ah\ absorption in \ion{N}{5}, the absence of \ion{C}{4} absorption in the low-state rules out a foreground origin of system B. As we show below, a plausible location is between the BLR and NLR, as for system \Ah.

\subsubsection{An additional system?}
An apparently significant dip appears at $\sim$~1530--1540~\AA\ in the low-state spectrum (noted by ``?'' in Fig.\ 1). We can rule out that this dip is produced by the excited \ion{Si}{2}$^{\ast}$~$\lambda1533.43$ since other excited absorption lines, in particular \ion{Si}{2}$^{\ast}$~$\lambda\lambda1264.77$, 1265.00, are not detected in the low-state spectrum. Alternatively, this may be a $v_{\rm shift}\simeq -3000$~\kms\ absorption system of \ion{C}{4}~$\lambda1548.19$, however, this identification is questionable as there is no clear feature at the expected position of \ion{C}{4}~$\lambda1550.77$. Also, none of the other ions shows a feature at this $v_{\rm shift}$. We suspect that this dip is an instrumental artifact which mostly affected the low 2 spectrum (Fig.\ 1), as this spectrum also shows apparent emission features at $\sim1530$ and 1540~\AA, which are not present in the low 1 spectrum.

\subsection{Absorbers inside the NLR}
As described in \S~3.3.2, the \ion{N}{5} and \ion{C}{4} absorption features can be reconstructed in both the low- and high-states assuming a fixed-profile absorber inside the NLR and outside the BLR (see Table 2 for the measured \nion). The assumption of a fixed-profile absorber was made for the sake of simplicity, as the dilution of the absorption by the strong NLR emission in the low-state does not allow a strong constraint on the low-state absorption profile. Photoionization modeling indicates that the large change in the ionizing continuum between the low- and high-states results in significant changes in the absorbing ionic columns, as well as the absorption profiles, even when the absorber properties, $R$, $n_{\rm e}$, and $\Sigma$ remain fixed. Can the low- and high-state spectra be reconstructed using a fixed properties absorber with variable ionization?

A consistent solution for a fixed absorber is found for an absorber located between the NLR and the BLR using ${\rm CF}=0.7$ and 1 for systems \Ah\ and B, respectively. The values of $\Sigma$ and $U$ which allow an acceptable reconstruction of the spectrum are $3\times10^{20}$~\cm\ and $0.2$ for system \Ah, and $10^{19}$~\cm\ and $0.02$ for system B ($U$ is $\sim5.5$ times lower during the low-state). These values correspond to $R\sim 10^{-6} - 10^{-2}$ and $10^{-5} - 10^{-1}$~pc for systems \Ah\ and B, respectively, for the range of $n_{\rm e}\sim10^5 - 10^{12}$~\cmth\ covered by our photoionization simulation grid. The range of $n_{\rm e}$ values could be much larger, making $R$ essentially unconstrained purely based on photoionization modeling. As we show in this work, rather tight constraints on $R$ can be achieved through the spectral reconstruction technique.

The absorber of both systems is optically-thin, having $b\sim100-200$~\kms. The estimated \nion\ at the low-state for \ion{N}{5}, \ion{C}{4}, and \ion{Si}{4} are: $1.0\times10^{16}$, $2.0\times10^{16}$, and $3.2\times10^{14}$~\cm, respectively, for system \Ah; and $4.0\times10^{13}$, $6.3\times10^{14}$, and $1.0\times10^{14}$~\cm\ for system B. The predicted \ion{Si}{4} column-density introduces a small emission feature in the reconstructed spectrum, which is consistent with noise (i.e.\ EW $\lesssim0.2$~\AA).\footnote{The photoionization model also predicts a significant \ion{C}{2} \nion\ for system B ($\sim 10^{14}$~\cm). Including this absorption in the reconstruction process of the \ion{C}{2} absorption-line feature, in addition to the low-ionization species absorber (system \Al), produces a small emission feature that is also consistent with the noise.} The best fit EW for the \ion{N}{5} and \ion{C}{4} narrow-line emission remain 8.5 and 75~\AA, respectively. The adopted EW for the \ion{C}{4} doublet is unusually high for AGN (e.g., Baskin \& Laor 2005, and references therein). It may be a by product of the large and rapid continuum variability in \ngc, which may produce an extended dust-free zone outside the BLR (Laor 2004, \S~3 there).

The adopted values of $\Sigma$ and $U$ imply a detectable intrinsic absorption by the \ion{O}{6} $\lambda\lambda$1031.93, 1037.62 doublet. The estimated \ion{O}{6} \nion\ is approximately $4\times10^{16}$~\cm\ for system \Ah\ at both flux states (corresponding to an EW $\sim2$~\AA\ for \ion{O}{6} $\lambda$1031.93), and $7\times10^{14}$ and $3\times10^{13}$~\cm\ for system B at the high- and low-states, respectively (${\rm EW}\sim0.7$ and $<0.01$~\AA, respectively). However, Crenshaw et~al.\ (2004) do not detect \ion{O}{6} absorption in a {\it FUSE} spectrum of \ngc, with an upper limit of $\sim0.3$~\AA\ for \ion{O}{6} $\lambda$1031.93, well below the values expected for system \Ah. However, a close examination of the {\it FUSE} spectrum (Crenshaw et~al.\ 2004) reveals no significant broad emission components for the \ion{O}{6}-doublet. This suggests that the object was in a low-state, where the spectrum is dominated by NLR emission which dilutes the absorption features and makes them unobservable, as seen for \ion{N}{5} at the low-state. We note in addition that the observed emission flux ratio of the \ion{O}{6}-doublet is 3:1, rather than the maximum ratio of 2:1 allowed by atomic physics. There is also an apparent absorption feature near \ion{O}{6} $\lambda$1037.62, but no corresponding feature near \ion{O}{6} $\lambda$1031.93, indicating that this is not due to \ion{O}{6}. Both effects suggest the presence of various systematic effects, either due to calibration, or due to contamination by significant stellar emission within the relatively large {\it FUSE} aperture, as suggested by Crenshaw et~al.\ (2004). Furthermore, Crenshaw et~al.\ also predict a detectable \ion{O}{6} absorption from the warm X-ray absorber (\S~5.3 there), which is not observed, possibly due to the reasons noted above.

Other combinations of $\Sigma$ and $U$ which yield larger \nion\ for \ion{N}{5} and \ion{C}{4}, do not allow an acceptable reconstruction of the low-state spectrum. Larger values of $\Sigma$, for the same $U$, result in a partially ionized region in the absorber, which produces stronger than observed absorption for \ion{Si}{2} and \ion{C}{2}, in particular in the low-state. Adopting $b\sim10$~\kms, and thus the upper limit of the \nion\ values, requires $\Sigma\gtrsim10^{24}$~\cm\ in order to produce large enough columns of C and N. This large value of $\Sigma$, which is untypical for gas found between the BLR and the NLR, also produces a prominent absorption by \ion{Si}{4} and \ion{C}{2} ($\sim1$ and 3~\AA, respectively) that is not observed here. Adopting ${\rm CF}=1$ for system \Ah\ produces a prominent emission feature in the \ion{N}{5} reconstruction in the low-state, which cannot be fitted with a narrow-line emission. Adopting ${\rm CF}=0.5$ for system B leads to \nion\ which cannot be reconstructed using the photoionization models. The N and C atoms in the above adopted fixed-absorber solution are mostly ionized beyond \ion{N}{5} and \ion{C}{4} in system \Ah\ in the high-state. A lower ionization solution, where \ion{N}{5} and \ion{C}{4} have not reached their maximal columns can be ruled out, as it will contain detectable column of low-ionization species, which are not observed. An ionization state where \ion{N}{5} and \ion{C}{4} have their maximal column can also be ruled out, because it produces detectable \ion{Si}{4} absorption during the high-state.\footnote{For a given $\Sigma$, \ion{Si}{4}, \ion{N}{5}, and \ion{C}{4} reach their maximal column at similar values of $U$ (especially \ion{Si}{4} and \ion{C}{4}).}

The system B absorber is less well constrained, but similar considerations can rule out $\Sigma$ larger than a few times $10^{19}$~\cm. It should be noted that unlike system \Ah, the \ion{N}{5} and \ion{C}{4} columns for system B during the high-state are closer to the maximal values. Constrains on the absorber metallicity and distance are discussed below in \S\S~5.2.1 and 5.2.2.

\subsubsection{Constraints on the absorber metallicity}
The absorber parameters were deduced assuming solar abundances, for the sake of simplicity. However, there are indications that the host galaxy metallicity (e.g.\ Roy et~al.\ 1996) and the metallicity of the NLR are also subsolar (Kraemer et~al.\ 1999). At subsolar metallicity the N/C abundance ratio is expected to scale as the metallicity (Hamann \& Ferland 1999, \S~6.4 there and references therein).\footnote{The predicted scaling with metallicity $Z$ of C/H, Si/H, and N/H is $Z$, $Z$, and $Z^2$, respectively.} We find that a N/C abundance ratio down to $\sim0.8$ times solar (i.e., 0.6 solar N/H) can still allow an acceptable fit to the observed absorption. Kraemer et~al.\ (1999) fit the emission lines of the NLR and BLR with a lower N/C abundance ratio of 0.4 solar, while assuming C/H and N/H abundance ratios of 1/2 and 1/6 solar, respectively. Such a low ratio cannot be clearly excluded by our modeling of the \ion{N}{5} and \ion{C}{4} absorption, however it implies a \ion{Si}{4} column-density of $10^{15}$~\cm\ for system \Ah\ in the low-state, and an absorption EW of $\sim2.5$~\AA\ (for \ion{Si}{4} $\lambda$1393.76), which is clearly not observed (the upper limit is 0.2~\AA). Despite this discrepancy we cannot take the absence of \ion{Si}{4} absorption as a robust evidence against the subsolar metallicity as adopted by Kraemer et~al.\ (1999), because Si may be heavily depleted into grains outside the BLR (e.g.\ Netzer \& Laor 1993).

The C/O ratio of 1/2 adopted by Kraemer et~al.\ (1999) is larger than predicted based on \ion{O}{3}] $\lambda1663$/\ion{C}{3}] $\lambda1909$ line ratio using Eq.\ 2 of Netzer (1997). Using the dereddened line ratios from Kraemer et~al.\ (1999, Table 1 there), Eq.\ 2 from Netzer, and assuming $T=15,000$~K one finds that ${\rm C/O}=0.16$. Because Kraemer et~al.\ scaled all the elements heavier than He relative to O, the C/H abundance ratio should be $5.4\times10^{-5}$. Thus, the N/C abundance ratio should be 0.4, which is similar to the solar abundance ratio (0.3). Using the revised subsolar abundances, one yields $\Sigma$ larger by a factor of $\sim6$ and similar values of $U$ and CF, as expected. However, \ion{Si}{4} remains over predicted in the low-state, because of over-abundance of Si relative to C and N by a factor of 6 relative to solar metallicity. It should be noted that the deduction of the location of the absorbers, i.e.\ between the BLR and the NLR, is not effected by the assumed absorber metallicity.

\subsubsection{Constrains on the absorber distance}
The distance of the absorbing systems can be constrained by the consistency requirement that the thickness of the absorber $d\sim\Sigma/n_{\rm e}$ is smaller than the distance, i.e. $d/R\ll 1$. This condition is fulfilled within the whole grid of photoionization models considered here (\S~4). The ionization timescale at $R<0.1$~pc is $<10$~days, which is shorter than the $\sim 100$ days between the two observations, and thus our assumption of photoionization equilibrium is valid. Peterson et~al.\ (2005) measured  $R_{\rm BLR}\sim 4\times10^{-5}$~pc for the \ion{C}{4} emitting region in a reverberation study of \ngc, and Kraemer et~al.\ (1999) estimated that ``INNER'' NLR, where most of the narrow \ion{C}{4} is produced, is located at $2.3\times 10^{-2}$~pc. This allows us to narrow down the range of values for $R$ from almost unconstrained (or $10^{-6} - 10^{-1}$~pc for our photoionization grid), to $4\times10^{-5}-2.3\times 10^{-2}$~pc implied by the spectral reconstruction technique which gives $R_{\rm BLR}<R<R_{\rm NLR}$.

The mass-loss rate through the outflowing systems can be estimated using $\dot{M}_{\rm out}=4\pi R \Sigma v_{\rm shift} m_{\rm p} C_{\rm g}$, where $m_{\rm p}$ is the proton mass and $C_{\rm g}$ is a global CF (e.g., Crenshaw et~al.\ 2003, \S~3.2.4 there). This yields $\dot{M}_{\rm out}= 3\times 10^{-7} - 1.7\times 10^{-4}~M_\sun\rm{~yr}^{-1}$ for system \Ah\ and $\sim 10$ times lower for system B, assuming $C_{\rm g}=1$. The bolometric luminosity of \ngc\ is about  $10^{40}$~erg~s$^{-1}$, which implies an accretion rate of $\dot{M}_{\rm in}=2\times 10^{-6}~M_\sun\rm{~yr}^{-1}$ assuming a 10\% efficiency. Thus, the requirement that $\dot{M}_{\rm in}\sim \dot{M}_{\rm out}$ also implies an absorber which is located between the BLR and the NLR.

\subsection{Comparison with earlier studies}
In a recent study, Crenshaw et~al.\ (2004) used high resolution UV spectra to investigate the UV absorption in \ngc, which was at a flux similar to the low-state. They report the detection of three absorption components, 1, 2, and G. Component G is attributed to the Galaxy. Component 2 is shifted by $\sim300$~\kms\ with respect to the Galaxy, and is attributed  to the ISM of the host galaxy. The velocity separation between components G and 2 is below our spectral resolution, and both components will be blended in our spectra. The weighted mean velocity-shift of components 2 and G is $\sim-220$~\kms\ based on the \ion{C}{2}~$\lambda1334.53$ absorption line (relative to the rest-frame used here)\footnote{Note that Crenshaw et~al.\ (2004) use $z=0.0012$, which is offset by $\sim-40$~\kms\ relative to the rest-frame used in this study.}. This velocity-shift is consistent with our adopted value for systems \Al\ and \Ah\ ($-250$~\kms). Crenshaw et~al.\ report the detection of components 2 and G in \ion{C}{2}, \ion{O}{1}, \ion{C}{4} and \ion{Si}{4}. The EWs measured for \ion{C}{2} and \ion{O}{1} in the current analysis are consistent with Crenshaw et~al.\ measurements within the uncertainties, where we sum over their EW measurements for components 2 and G. The absorption by \ion{C}{4} and \ion{Si}{4} detected by Crenshaw et~al.\ can be attributed to the high-ionization absorber (system \Ah). These absorption features are narrow ($\sim 100$~\kms) and are thus detectable in the high resolution UV spectra, despite the strong NLR contribution at the low-state, which does not dilute the absorption as it does in our lower resolution spectra. The EW of 0.59~\AA\ predicted here for \ion{C}{4}~$\lambda1548.20$ in the low-state is consistent with Crenshaw et~al.\ \ion{C}{4} measurement of 0.77~\AA\ (for components G+2), although all the \ion{C}{4} absorption in our model is attributed to the intrinsic absorption system. The estimated \ion{Si}{4} EW in the current study, 0.2~\AA\ for \ion{Si}{4}~$\lambda1393.76$, is much smaller than measured by Crenshaw et~al., 1.2 plus 0.4~\AA\ for components G and 2, respectively. Most of the discrepancy is caused by component G, which has a negative flux in a significant part of its absorption trough (see Fig.\ 2 there), undermining the accuracy of the measured EW. Crenshaw et~al.\ attribute component 2 to the ISM of the host galaxy, which is consistent with the interpretation adopted here for the low-ionization species absorber (system \Al).

Component 1 was detected by Crenshaw et~al.\ (2004) in \ion{C}{4}~$\lambda1548.20$ at $v_{\rm shift}=-730$~\kms\ (relative to the rest-frame used here), and this component is consistent with our system B. The \ion{C}{4}~$\lambda1548.20$ absorption EW estimated by Crenshaw et~al.\ is larger than the predicted value by the current study (0.8 versus $\sim0.2$~\AA), but it is probably within the uncertainty given the low S/N of the high resolution spectra. To summarize, the present results for the low-state are mostly consistent with the work of Crenshaw et~al.\ (2004).

Several X-ray studies of \ngc\ find an intrinsic X-ray absorber. Iwasawa et~al.\ (2000) estimate $\Sigma\sim2-10\times10^{22}$~cm$^{-2}$ for a warm absorber and $\sim2-5\times10^{21}$~cm$^{-2}$ for a cold absorber, based on {\it ASCA} observation. Shih et~al.\ (2003) and Moran et~al.\ (2005) deduce similar values based on {\it ASCA} and {\it Chandra} observations, respectively. Thus, $\Sigma\sim3\times10^{20}$~cm$^{-2}$ measured for system \Ah\ is inconsistent with the suggested cold X-ray absorber (system B has an even smaller value of $\Sigma$). As mentioned above, larger values of $\Sigma$ for systems \Ah\ and B can be ruled out, because they predict  additional absorption by low-ionization species which is not detected here. Crenshaw et~al.\ (2004) also fail to detect in {\it FUSE} spectra \ion{H}{1} and \ion{O}{6} absorption predicted by photoionization models for the X-ray absorber from Shih et~al.\ (2003), although their conclusion is inconclusive. It should be mentioned in passing that Moran et~al.\ (1999) estimate $\Sigma\sim2\times10^{20}$~cm$^{-2}$ for an X-ray absorber based on a low-count {\it ROSAT} observation. They attribute this absorber to the Galaxy based on a similarity of the Galactic $\Sigma$. Based on the current study, one can also place the X-ray absorber observed by them in the absorption system \Ah. The warm $\Sigma\sim2-10\times10^{22}$~cm$^{-2}$ absorber is most likely undetectable in the UV as N and C are too highly ionized.

It is interesting to note that the value of $\Sigma=3\times10^{20}$~\cm\ obtained here for system \Ah\ is close to the value for the broad-line emitting gas, $\Sigma\sim7\times10^{20}$~\cm, deduced by Kraemer et~al.\ (1999). Adopting $R_{\rm BLR}\sim4\times10^{-5}$~pc from Peterson et~al.\ (2005) for system \Ah, yields $n_{\rm e}\sim(1-10)\times10^{10}$~\cmth\ which is also consistent with the Kraemer et~al.\ deduced density for the BLR ($3\times10^{10}$~\cmth).  The column-density estimated by Kraemer et~al.\ (1999) for the NLR, $\Sigma_{\rm NLR}\sim10^{21}$~\cm\, is significantly larger than our estimates for systems \Ah\ and B. Thus, system \Ah\ may be very close to the BLR, and it could be produced by BLR gas which happens to cross our line of sight to the nucleus.

The current study is one of only a few, which place a geometrical constraint on the location of a UV absorber without direct knowledge of its exact distance from the nucleus based on $n_{\rm e}$ (e.g., Kraemer et~al.\ 2001), or knowledge on its location based on CF considerations (e.g., Gabel et~al.\ 2005a). Gabel et~al.\ (2005b) reach the conclusion that the absorber in NGC~3783 is located interior to the NLR by comparing the observed emission features in the high- and low-state spectra. Arav, Korista, \& de~Kool (2002) reach similar conclusion for NGC~5548 by modeling the continuum, BLR, and NLR emission and comparing it to the observed one. Arav et~al.\ (1999) place the absorber in PG~1603+3002 interior to the BLR using similar considerations. A portion of the absorbing gas detected in NGC~4151 is placed outside the ``intermediate'' line region by Kraemer et~al.\ (2006) based on decomposition of the \ion{C}{4} profile. It should be noted that we do not assume nor constrain the location of the absorber prior to carrying out the reconstruction procedure (i.e., measurement of the absorption), but rather allow for all possible locations, which are then scrutinized using atomic-physics and photoionization considerations. This procedure also does not rely on a decomposition of emission features, which is often not well constrained. The main shortcoming of our approach is that it cannot be used in single epoch studies (unlike the decomposition methods).

\section{CONCLUSIONS}
We report the detection of variable UV absorption lines towards the nucleus of the least luminous Seyfert 1 galaxy, \ngc, based on observations with \hst\ STIS that were carried out as part of a reverberation-mapping program (Peterson et~al.\ 2005). Two sets of observations were obtained $\sim3$~month apart. \ngc\ was in a low-state during the first set of observations, and in a high-state in the second set, where the continuum flux increased by factors of 4--7. Low-ionization lines of \ion{O}{1}, \ion{N}{1}, \ion{Si}{2}, \ion{C}{2}, and \ion{Fe}{2}, are present in the low-state (April 2004) at a velocity $v_{\rm shift}=-250$~\kms\ (system \Al), and additional high-ionization lines of \ion{C}{4} and \ion{N}{5} appear in the high-state (July 2004) at $v_{\rm shift}=-250$~\kms\ (system \Ah) and at $v_{\rm shift}=-840$~\kms\ (system B). In order to measure the absorption-line properties we use a spectrum reconstruction technique, which does not assume an unabsorbed spectral shape a priori. Finally, we use photoionization models in order to constrain the physical location of the absorber and its parameters.

We reach the following conclusions:
\begin{enumerate}
\item The system \Ah\ \ion{N}{5} absorption EW in the high-state is $\ga 10$ times larger than the  \ion{C}{4} absorption EW. Photoionization models indicate that such a high ratio can be obtained only if the N/C abundance ratio is significantly above solar. However, the metallicity of the host galaxy, the NLR and the BLR are all significantly subsolar, which strongly argues against absorption by a high metallicity system.
\item The disappearance of the \ion{N}{5} absorption in the low-state cannot be explained by a drop in the ionization state of a foreground absorbing gas, as the expected absorption by lower-ionization species, in particular \ion{Si}{4}, does not show up. It is also unlikely that \ion{N}{4} could be photoionized to \ion{N}{5} in the three months between the low- and high-states. This excludes a photoionized foreground absorber regardless of metallicity.
\item The high N/C absorption EW ratio, and the disappearance of the \ion{N}{5} absorption in the low-state, can both be explained by a fixed density and column-density absorber located between the BLR and the NLR. The apparent disappearance of the \ion{N}{5} absorption in the low-state results from the large drop in the absorbed continuum and BLR emission, and the strong dilution by the NLR emission which fills up the \ion{N}{5} absorption troughs.
\item An absorber located inside the BLR is ruled out, as absorption of the continuum only cannot produce a deep enough feature for the two absorption systems seen in \ion{C}{4}.
\item The favored values of $\Sigma\sim3\times10^{20}$~\cm\ and $U\sim0.2$ for system \Ah\ are similar to those deduced by Kraemer et~al.\ (1999) for the BLR emitting gas, suggesting this system may be produced by an outer BLR gas along the line of sight.
\item The UV absorption systems are probably not related to the X-ray absorbers, which require column-densities one to three magnitudes larger than estimated here (e.g, Iwasawa et~al.\ 2000).
\item The low-ionization species absorber (i.e.\ system \Al) does not vary between the two flux-states. It also does not show absorption by metastable level, in particular in the \ion{Si}{2}$^\ast$ $\lambda\lambda1264.74, 1265.02$ lines. This implies that the density of the absorbing gas is lower than $\sim1.8\times10^3$~cm$^{-3}$, and it must thus be located outside the NLR of \ngc, or possibly in the ISM of the host or in the Galactic ISM (or both; see Crenshaw et~al.\ 2004).
\end{enumerate}

Apparent changes in the absorption strength are commonly interpreted as changes in the absorbing medium (e.g.\ ``moving clouds''). However, as we have shown here, such changes may also result from changes in the illumination pattern behind a fixed absorber. If the absorber lies behind some radiation sources (the NLR here), and in front of others (the BLR plus continuum here), then large changes in the relative strength of the absorbed versus unabsorbed sources can produce large changes in the absorption features, even when the absorber is not variable. The study of the UV absorption line variability in objects showing large continuum variations can thus provide new constraints on the location of the absorber, and thus break the large degeneracy in the absorber distance determination based on single epoch spectra. A simple prediction of our fixed absorber model is that the strength of the absorption features should be strongly correlated with luminosity, whereas in the ``moving clouds'' model no such correlation is expected.

\acknowledgments
We would like to thank N.\ Arav for helpful discussions, G.\ Ferland for making {\sc cloudy} publicly available, and the anonymous referee for comments and suggestions. This research has made use of the NASA/IPAC Extragalactic Database (NED), which is operated by the Jet Propulsion Laboratory, California Institute of Technology, under contract with the National Aeronautics and Space Administration. This research has made use of the STIS electronic manual, which is operated by the Space Telescope Science Institute, under contract with the National Aeronautics and Space Administration. This research was supported by The Israel Science Foundation (grant 1030/04), and by a grant from the Norman and Helen Asher Space Research Institute.


\clearpage
\begin{deluxetable}{@{}l*{1}{r}l*{6}{r}@{}}
 \tablecaption{The estimated \nion\ and EW for an absorbing foreground screen outside the NLR.\label{measN}}

\tablehead{\multicolumn{1}{c}{Ion} & \multicolumn{1}{c}{$\lambda_{\rm ion}$} &
\multicolumn{1}{c}{State\tablenotemark{a}} &\multicolumn{4}{c}{${\rm CF}=1$} &
\multicolumn{2}{c}{${\rm CF}=0.75$\tablenotemark{b}}\\
& & & \multicolumn{2}{c}{Lower-limit\tablenotemark{c}} &
\multicolumn{2}{c}{Upper-limit\tablenotemark{d}} & &
\\ & & & \multicolumn{1}{c}{$N_{\rm ion}$} & \multicolumn{1}{c}{EW} &
\multicolumn{1}{c}{$N_{\rm ion}$} & \multicolumn{1}{c}{EW} &
\multicolumn{1}{c}{$N_{\rm ion}$} & \multicolumn{1}{c}{EW}\\
\multicolumn{1}{c}{(1)} & \multicolumn{1}{c}{(2)} & \multicolumn{1}{c}{(3)} &
\multicolumn{1}{c}{(4)} & \multicolumn{1}{c}{(5)} & \multicolumn{1}{c}{(6)} &
\multicolumn{1}{c}{(7)} & \multicolumn{1}{c}{(8)} & \multicolumn{1}{c}{(9)}}

\startdata
\multicolumn{9}{c}{\textbf{System \Al} ($v_{\rm shift}=-250$~\kms)}\\
\ion{C}{1}\tablenotemark{e}& 1560.31 & low &6.0E+14 &   0.85    &   2.0E+19 &   0.84    &   8.5E+14 &   0.83    \\
& & high &$<$5.0E+13 &   $<$0.09    &   $<$5.0E+13 &   $<$0.07    &   $<$5.0E+13 &   $<$0.06    \\
\ion{C}{2}\tablenotemark{e}&   1334.53 & low &  8.0E+14 &   1.13    &   1.5E+19 &   1.18    &   1.1E+15 &   1.04    \\
& & high & 1.4E+15 &   1.58    &   3.5E+19 &   1.79    &   2.5E+15 &   1.53    \\
\ion{N}{1}&   1199.97 & &  9.0E+14 &   1.56    &   1.5E+19 &   1.78    &   2.0E+15 &   1.56    \\
\ion{O}{1}&   1302.17 & &  2.0E+15 &   1.05    &   3.5E+19 &   1.23    &   3.5E+15 &   1.11    \\
\ion{Si}{1}\tablenotemark{e}&   1562.00 & low &  2.0E+13 &   0.16    &   1.0E+15 &   0.24    &   3.0E+13 &   0.17    \\
& & high & $<$1.0E+13 &   $<$0.08    &   $<$1.0E+13 &   $<$0.07    &   $<$1.0E+13 &   $<$0.06    \\
\ion{Si}{2}&   1190.42 & &  3.0E+14 &   0.74    &   1.0E+18 &   0.56    &   7.0E+14 &   0.98    \\
&   1193.29 &  & 3.0E+14 &   1.20    &   1.0E+18 &   1.49    &   7.0E+14 &   1.35    \\
&   1260.42 &  & 3.0E+14 &   1.84    &   1.0E+18 &   2.24    &   7.0E+14 &   1.77    \\
&   1304.37 &  & 3.0E+14 &   0.57    &   1.0E+18 &   0.52    &   7.0E+14 &   0.82    \\
&   1526.71 &  & 3.0E+14 &   1.08    &   1.0E+18 &   0.90    &   7.0E+14 &   1.38    \\
\ion{Fe}{2} & 1608.45 & & 8.0E+14 & 0.91 & 6.0E+18 & 0.73 & 1.3E+15 & 0.98 \\
\\
\multicolumn{9}{c}{\textbf{System \Ah} ($v_{\rm shift}=-250$~\kms)}\\
\ion{C}{4}&   1548.19 & &  $<$5.0E+13 &   $<$0.19    &   $<$1.0E+15 &   $<$0.22    &   $<$5.0E+13 &   $<$0.15    \\
&   1550.77 &  & $<$5.0E+13 &   $<$0.10    &   $<$1.0E+15 &   $<$0.20    &   $<$5.0E+13 &   $<$0.07    \\
\ion{N}{5}&   1238.82 & low & $<$5.0E+13 &   $<$0.10    &   $<$1.0E+14 &   $<$0.10    &   $<$5.0E+13 &   $<$0.08    \\
& & high & 1.4E+15 &   1.57    &   1.8E+19 &   1.46    &   2.5E+15 &   1.48    \\
&   1242.80 & low & $<$5.0E+13 &   $<$0.05    &   $<$1.0E+14 &   $<$0.07    &   $<$5.0E+13 &   $<$0.04    \\
& & high & 1.4E+15 &   1.05    &   1.8E+19 &   1.04    &   2.5E+15 &   1.11    \\
\ion{Si}{4}&   1393.76 &  & $<$2.0E+13 &   $<$0.17    &   $<$5.0E+14 &   $<$0.20    &   $<$2.0E+13 &   $<$0.13    \\
&   1402.77 &  & $<$2.0E+13 &   $<$0.09    &   $<$5.0E+14 &   $<$0.19    &  $<$2.0E+13 &   $<$0.07    \\
\\
\multicolumn{9}{c}{\textbf{System B} ($v_{\rm shift}=-840$~\kms)}\\
\ion{C}{4}&   1548.19 & low&   $<$5.0E+13 &   $<$0.19    &   $<$5.0E+14 &   $<$0.20    &   $<$5.0E+13 &   $<$0.15    \\
& & high & 2.5E+14 &   0.83    &   2.0E+18 &   0.77    &   4.0E+14 &   0.89    \\
&   1550.77 & low&   $<$5.0E+13 &   $<$0.10    &   $<$5.0E+14 &   $<$0.18    &   $<$5.0E+13 &   $<$0.07    \\
& & high & 2.5E+14 &   0.46    &   2.0E+18 &   0.59    &   4.0E+14 &   0.52    \\
\ion{N}{5}&   1238.82 & low &   $<$5.0E+13 &   $<$0.10    &   $<$1.0E+14 &   $<$0.10    &   $<$5.0E+13 &   $<$0.08    \\
& & high & 1.5E+14 &   0.30    &   1.0E+18 &   0.41    &   2.0E+14 &   0.29    \\
&   1242.80 & low &   $<$5.0E+13 &   $<$0.05    &   $<$1.0E+14 &   $<$0.07    &   $<$5.0E+13 &   $<$0.04    \\
& & high & 1.5E+14 &   0.15    &   1.0E+18 &   0.34    &   2.0E+14 &   0.15    \\
\ion{Si}{4}&   1393.76 & &   $<$2.0E+13 &   $<$0.17    &   $<$5.0E+14 &   $<$0.20    &   $<$2.0E+13 &   $<$0.13    \\
&   1402.77 &   &   $<$2.0E+13 &   $<$0.09    &   $<$5.0E+14 &   $<$0.19 &
$<$2.0E+13 &  $<$0.07\\
\enddata

\tablenotetext{a}{The flux state of \ngc: low -- visit 1 and visit 2 mean spectrum; high -- visit 3 mean spectrum; (blank) -- the tabulated measurements are consistent for both flux states.}
\tablenotetext{b}{The lowest acceptable value of CF. Using $b=180$~\kms. A lower-limit of \nion. No upper-limit could be measured while assuming partial covering (see text).}
\tablenotetext{c}{Using $b=180$~\kms.}
\tablenotetext{d}{Using $b=10$~\kms.}
\tablenotetext{e}{The discrepancy between the measurements for the two flux states is probably an artifact of the spectra acquisition by the \hst\ STIS (see \S~5.1.1).}
\end{deluxetable}

\clearpage
\begin{deluxetable}{@{}l*{1}{r}l*{7}{r}@{}}
 \tablecaption{The estimated \nion\ and EW for the high-state, for an absorber between the BLR and the NLR.\tablenotemark{a}\label{measNBLR}}

\tablehead{\multicolumn{1}{c}{Ion} & \multicolumn{1}{c}{$\lambda_{\rm ion}$} &
\multicolumn{4}{c}{${\rm CF}=1$} &
\multicolumn{2}{c}{${\rm CF}<1$\tablenotemark{b}}\\
& & \multicolumn{2}{c}{Lower-limit\tablenotemark{c}} &
\multicolumn{2}{c}{Upper-limit\tablenotemark{d}} &
\multicolumn{2}{c}{}\\
& & \multicolumn{1}{c}{$N_{\rm ion}$} & \multicolumn{1}{c}{EW} &
\multicolumn{1}{c}{$N_{\rm ion}$} & \multicolumn{1}{c}{EW} &
\multicolumn{1}{c}{$N_{\rm ion}$} & \multicolumn{1}{c}{EW}\\
\multicolumn{1}{c}{(1)} & \multicolumn{1}{c}{(2)} & \multicolumn{1}{c}{(3)} &
\multicolumn{1}{c}{(4)} & \multicolumn{1}{c}{(5)} & \multicolumn{1}{c}{(6)} &
\multicolumn{1}{c}{(7)} & \multicolumn{1}{c}{(8)}}

\startdata
\multicolumn{8}{c}{\textbf{System \Ah}}\\
\multicolumn{6}{c}{}&\multicolumn{2}{c}{${\rm CF}=0.7$}\\
\ion{C}{4} & 1548.19  & 6.0E+14 & 1.55 & 8.0E+18 & 1.49 & 1.2E+15 & 1.54\\
& 1550.77  & 6.0E+14 & 0.96 & 8.0E+18 & 1.06 & 1.2E+15 & 1.09\\
\ion{N}{5}& 1238.82  & 1.9E+15 & 1.79 & 3.2E+19 & 1.94 & 5.0E+15 & 1.66\\
& 1242.80 &  1.9E+15 & 1.28 & 3.2E+19 & 1.38 & 5.0E+15 & 1.39\\\\
\multicolumn{8}{c}{\textbf{System B}}\\
\multicolumn{6}{c}{}&\multicolumn{2}{c}{${\rm CF}=0.5$}\\
\ion{C}{4} & 1548.19 & 5.0E+14 & 1.38 & 7.0E+18 & 1.39 & 2.5E+15 & 1.39\\
& 1550.77  & 5.0E+14 & 0.83 & 7.0E+18 & 1.00 & 2.5E+15 & 1.12\\
\ion{N}{5}& 1238.82  & 1.8E+14 & 0.35 & 1.0E+18 & 0.41 & 5.0E+14 & 0.41\\
& 1242.80 & 1.8E+14 & 0.18 & 1.0E+18 & 0.34 & 5.0E+14 & 0.23\\
\enddata

\tablenotetext{a}{The quantities are estimated from the high$-$low spectrum i.e, high-state spectrum with subtracted NLR emission.}
\tablenotetext{b}{The lowest acceptable value of CF. Using $b=180$~\kms. A lower-limit of \nion. No upper-limit could be measured while assuming partial covering. The \nion\ values can be modified by a factor of $\sim3$ without altering the reconstructed spectrum significantly.}
\tablenotetext{c}{Using $b=180$~\kms.}
\tablenotetext{d}{Using $b=10$~\kms.}
\end{deluxetable}

\clearpage
\begin{deluxetable}{@{}l*{1}{r}l*{7}{r}@{}}
 \tablecaption{The estimated \nion\ and EW for the high-state, for an absorber between the BLR and the continuum source.\tablenotemark{a}\label{measNCont}}

\tablehead{\multicolumn{1}{c}{Ion} & \multicolumn{1}{c}{$\lambda_{\rm ion}$} &
\multicolumn{4}{c}{${\rm CF}=1$} &
\multicolumn{2}{c}{${\rm CF}<1$\tablenotemark{b}}\\
& & \multicolumn{2}{c}{Lower-limit\tablenotemark{c}} &
\multicolumn{2}{c}{Upper-limit\tablenotemark{d}} &\\
& & \multicolumn{1}{c}{$N_{\rm ion}$} & \multicolumn{1}{c}{EW} &
\multicolumn{1}{c}{$N_{\rm ion}$} & \multicolumn{1}{c}{EW} &
\multicolumn{1}{c}{$N_{\rm ion}$} & \multicolumn{1}{c}{EW}\\
\multicolumn{1}{c}{(1)} & \multicolumn{1}{c}{(2)} & \multicolumn{1}{c}{(3)} &
\multicolumn{1}{c}{(4)} & \multicolumn{1}{c}{(5)} & \multicolumn{1}{c}{(6)} &
\multicolumn{1}{c}{(7)} & \multicolumn{1}{c}{(8)}}

\startdata
\multicolumn{8}{c}{\textbf{System \Ah}}\\
\multicolumn{6}{c}{}&\multicolumn{2}{c}{${\rm CF}=0.8$}\\
\ion{C}{4} & & \multicolumn{6}{c}{cannot be reconstructed}\\
\ion{N}{5}& 1238.82 &  1.1E+16 & 2.73 & 9.5E+19 & 3.32 & 2.0E+16 & 2.37\\
& 1242.80 &  1.1E+16 & 2.42 & 9.5E+19 & 2.36 & 2.0E+16 & 2.16\\\\
\multicolumn{8}{c}{\textbf{System B}}\\
\multicolumn{6}{c}{}&\multicolumn{2}{c}{${\rm CF}=0.5$}\\
\ion{C}{4} & & \multicolumn{6}{c}{cannot be reconstructed}\\
\ion{N}{5}& 1238.82  & 6.0E+14 & 0.94 & 4.0E+18 & 0.71 & 1.2E+15 & 0.72\\
& 1242.80 & 6.0E+14 & 0.55 & 4.0E+18 & 0.53 & 1.2E+15 & 0.47
\enddata

\tablenotetext{a}{The quantities are estimated from the high$-$low spectrum i.e, high-state spectrum with subtracted NLR emission.}
\tablenotetext{b}{The lowest acceptable value of CF. Using $b=180$~\kms. A lower-limit of \nion. No upper-limit could be measured while assuming partial covering. The \nion\ values can be modified by a factor of $\sim3$ without altering the reconstructed spectrum significantly.}
\tablenotetext{c}{Using $b=180$~\kms.}
\tablenotetext{d}{Using $b=10$~\kms.}
\end{deluxetable}

\clearpage
\begin{deluxetable}{@{}l*{1}r*{1}{l}@{}}
  \tablecaption{The spectral energy distribution adopted for the high-state.\tablenotemark{a}\label{SEDhigh}}

  \tablehead{\multicolumn{1}{c}{Energy range} & \multicolumn{1}{c}{$\alpha$\tablenotemark{b}} & \multicolumn{1}{c}{References}\\ \multicolumn{1}{c}{(eV)}&&}

  \startdata
  1 -- 2 & $0.85$ & Laor (2006)\\
  2 -- 6 & $0.90$ & ---\tablenotemark{c}\\
  6 -- 10.6 & $0.94$ & this paper\tablenotemark{d}\\
  10.6 -- $1\times10^3$ & $1.19$ & Iwasawa et~al.\ (2000) \& this paper\tablenotemark{e}\\
  $(1 - 40)\times10^3$ & $0.72$ & Iwasawa et~al.\ (2000)
  \enddata

  \tablenotetext{a}{A cutoff is assumed above 40~keV.}
  \tablenotetext{b}{$f_{\nu}\propto\nu^{-\alpha}$.}
  \tablenotetext{c}{The slope is approximately the mean value between 0.85 and 0.94.}
  \tablenotetext{d}{The slope is measured using the current spectrum between 1700 and 1170~\AA, and is extrapolated to 6~eV.}
  \tablenotetext{e}{The slope is measured using the flux at 1170~\AA\ of the current spectrum and the flux at 1~keV adopted from Iwasawa et~al.\ (2000, \S~6.2 there).}
\end{deluxetable}

\clearpage
\begin{deluxetable}{@{}l*{1}r*{1}{l}@{}}
  \tablecaption{The spectral energy distribution adopted for the low-state.\tablenotemark{a}\label{SEDlow}}

  \tablehead{\multicolumn{1}{c}{Energy range} & \multicolumn{1}{c}{$\alpha$\tablenotemark{b}} & \multicolumn{1}{c}{References}\\ \multicolumn{1}{c}{(eV)}&&}

  \startdata
  1 -- 7.3 & $2.26$ & Desroches et~al.\ (2006) \& this paper\tablenotemark{c}\\
  7.3 -- 10.6 & $2.36$ & this paper\\
  10.6 -- $1\times10^3$ & $1.37$ & Moran et~al.\ (1999) \& this paper\tablenotemark{d}\\
  $(1 - 40)\times10^3$ & $-0.39$ & Moran et~al. (2005)
  \enddata

  \tablenotetext{a}{A cutoff is assumed above 40~keV.}
  \tablenotetext{b}{$f_{\nu}\propto\nu^{-\alpha}$.}
  \tablenotetext{c}{The slope is calculated using the flux at 1700~\AA\ of the current low-state, and the flux at {\it V} band, which was measured simultaneously with the current low-state UV observations by Desroches et~al.\ (2006). This slope is then extrapolated to 1~eV.}
  \tablenotetext{d}{The slope is calculated using the flux at 1170~\AA\ of the current low-state and the flux at 1~keV reported by Moran et~al.\ (1999, Fig.\ 4 there). Note that Moran et~al.\ report only the measured luminosity. We transfer it to flux units using the ratio between the luminosities at 1~keV reported by Moran et~al.\ and Iwasawa et~al.\ (2000), multiplied by the flux at 1~keV reported in the latter study.}
\end{deluxetable}

\clearpage
\begin{figure}
\begin{center}
\includegraphics[width=12.0cm]{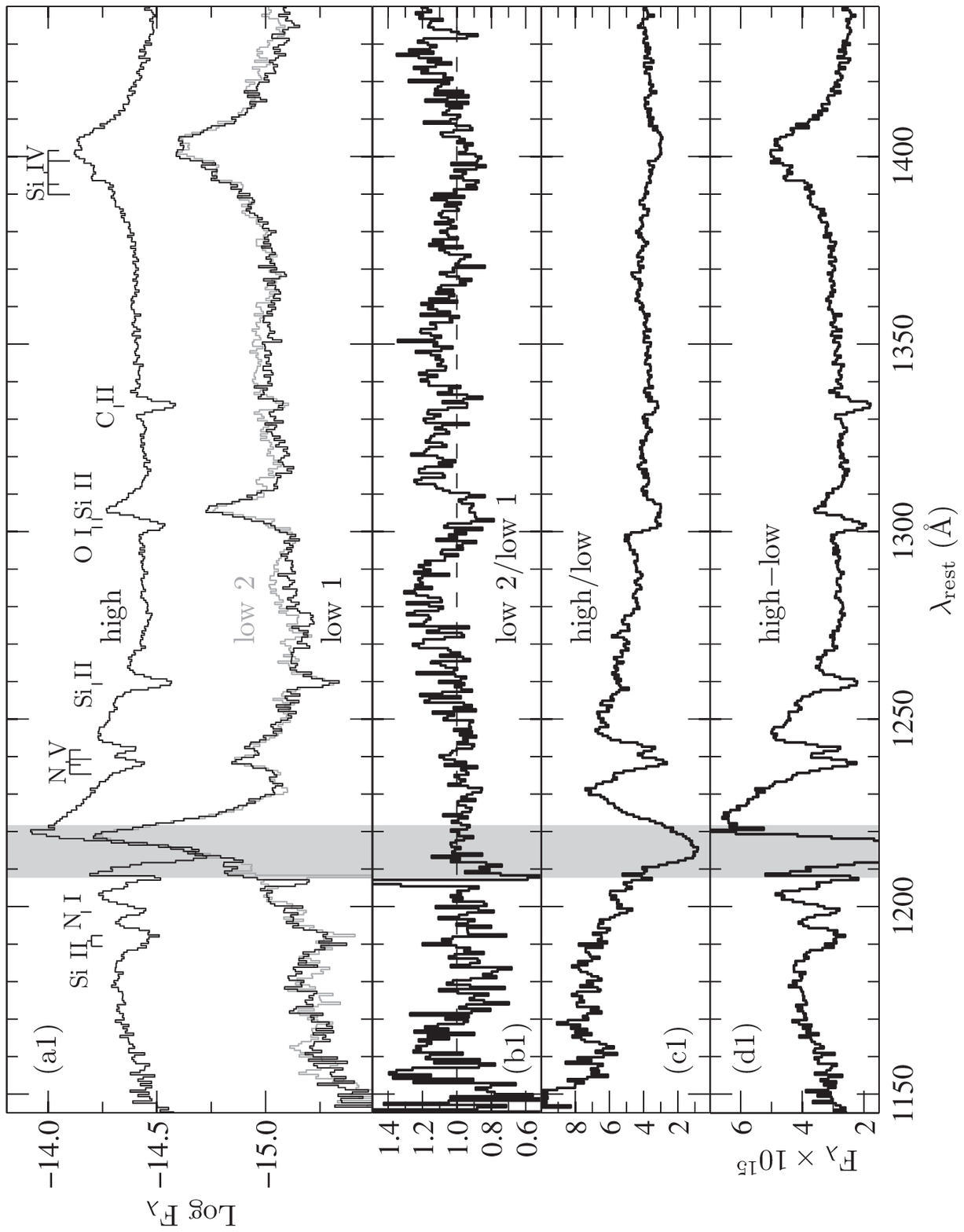}
\caption{The mean \hst\ STIS spectrum of the two low-states and the high-state of \ngc, corrected for Galactic extinction [panels (a1,a2)]. The ratio between the two low-states spectra [panels (b1,b2)], the ratio between the high-state spectrum and the mean low-state spectrum [panels (c1,c2)] and the difference between the high- and mean low-state spectra [panels (d1,d2)]. The grayed area indicates the wavelengths range of 1207.7--1221.7~\AA\ which is significantly affected by the Geocoronal \Lya\ emission and not included in the current analysis. The lines in panel (a) indicate the expected wavelength of absorption systems \Al\ and \Ah\ (short line) and B (long line), relative to the rest-frame of \ngc\ (see text). The question mark indicates an additional possible absorption system. Note the large continuum increase (factor of 4--7) between the low- and high-state [panels (c1,c2)]. Note the strong \ion{N}{5} absorption, but very weak \ion{C}{4} absorption in the high-state, and the disappearance of the \ion{N}{5} absorption in the low-states.}
\end{center}
\end{figure}

\clearpage
\begin{center}
\includegraphics[width=12.0cm]{f1b.eps}
\\Fig.~1 --- Continued.
\end{center}

\clearpage
\begin{figure}
\begin{center}
\includegraphics[angle=-90,width=12.7cm]{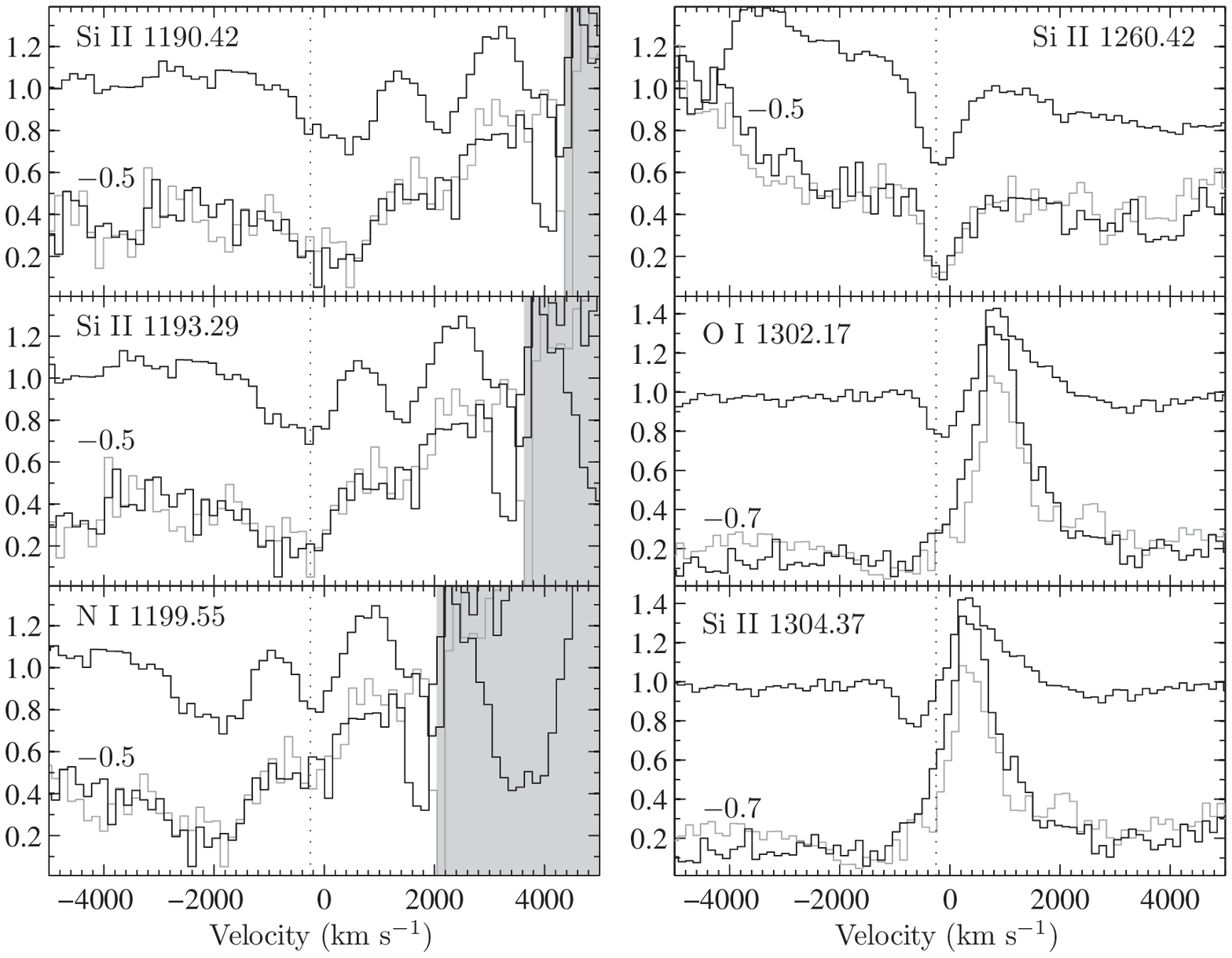}
\includegraphics[angle=-90,width=12.7cm]{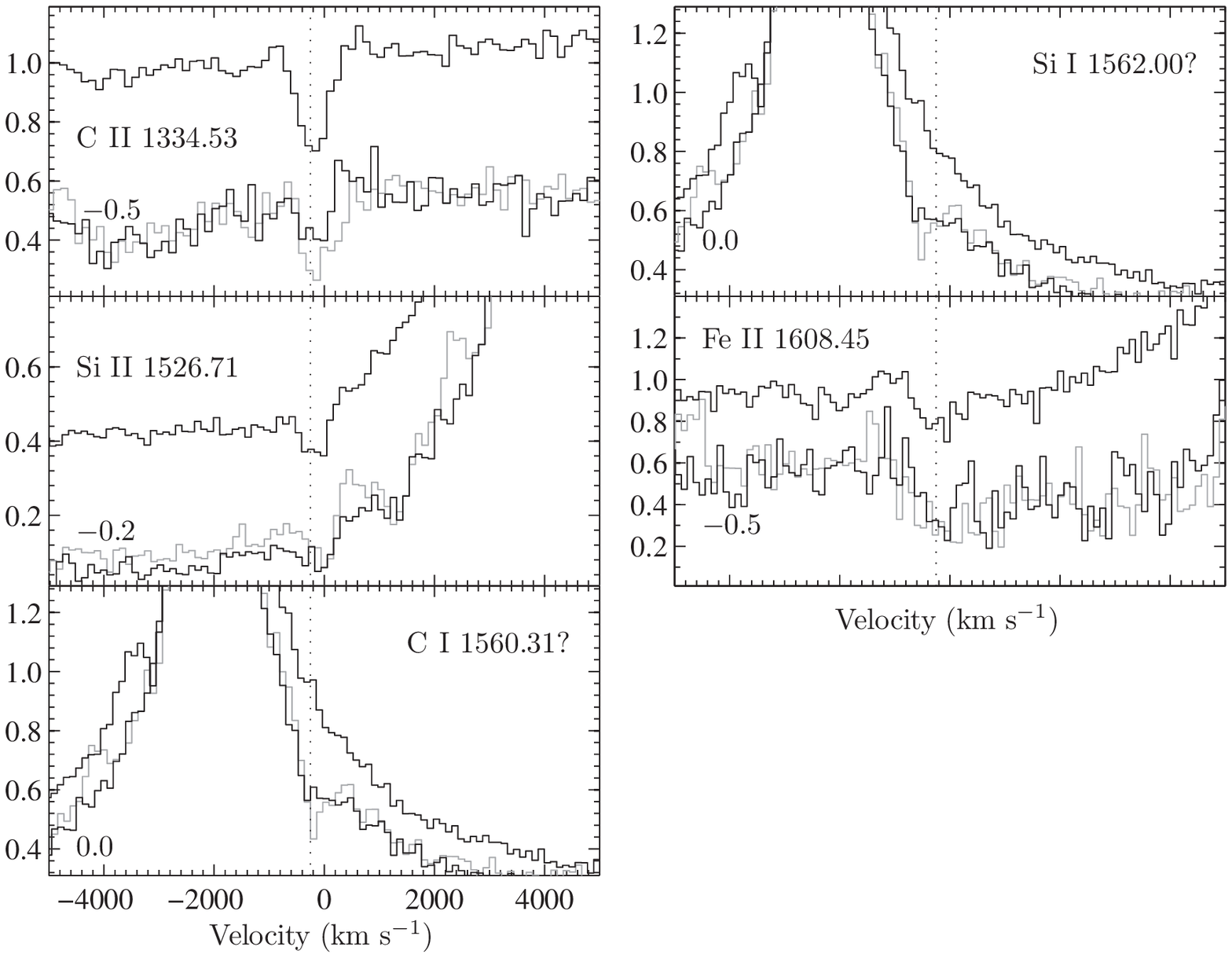}
\caption{The absorption-line profiles for the low-ionization species. The rest wavelength used to set the velocity scale is indicated in each panel. The spectra were normalized by the mean flux in the $\pm5000$~\kms\ neighborhood of the line center. The velocity-shift of $-250$~\kms\ of the absorption system \Al\ is indicated by a dashed line. The low 1 and low 2 spectra are shifted by a factor indicated in each panel. Note the general similarity between the absorption-line profiles of the low 1 and low 2 states.}
\end{center}
\end{figure}

\clearpage
\begin{figure}
\begin{center}
\includegraphics[angle=-90,width=12.7cm]{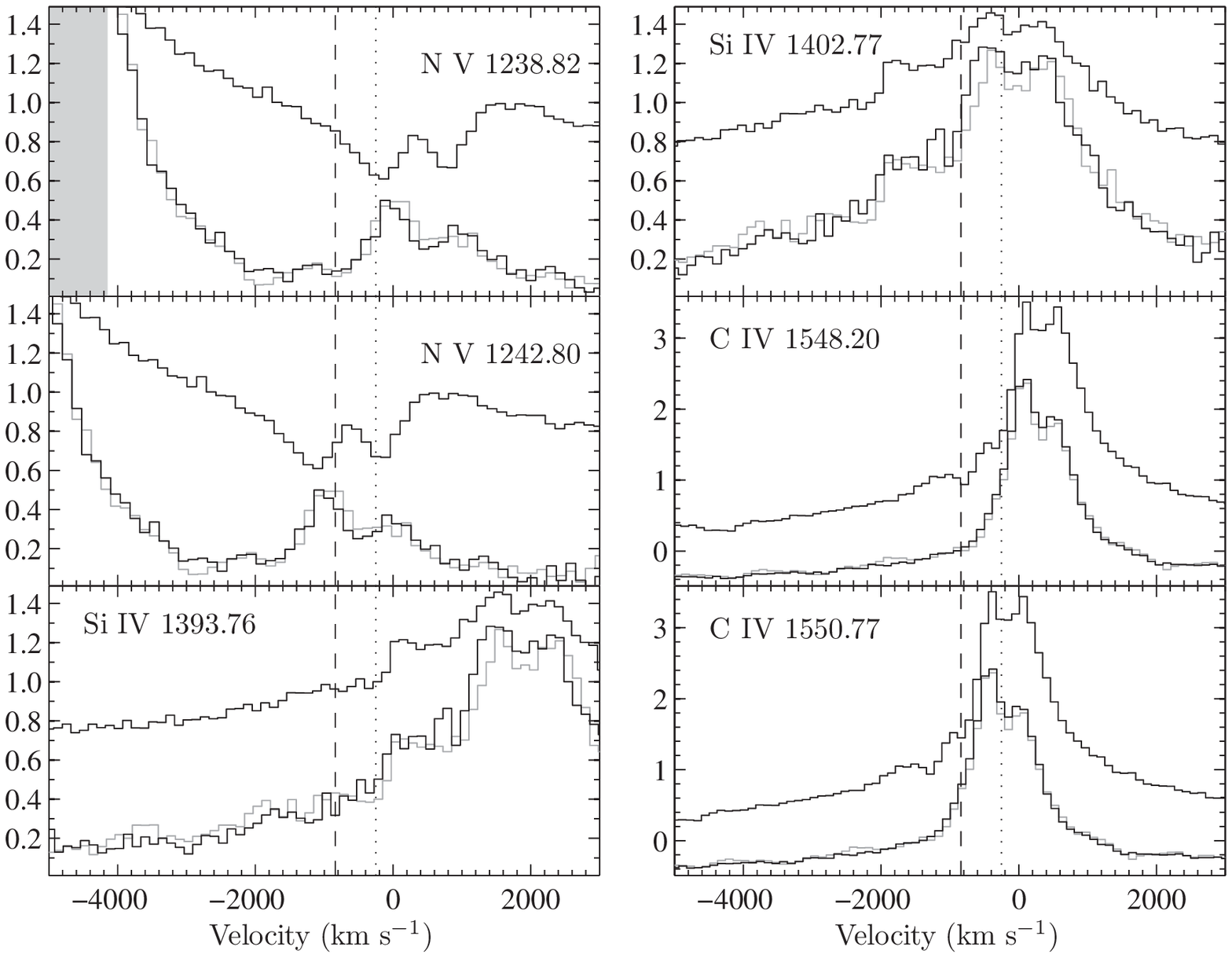}
\caption{The absorption-line profiles for the high-ionization species. The same normalization procedure and notation as in Fig.\ 2 are used. The velocity-shifts of the absorption systems \Ah\ (dotted) and B (dashed) are indicated ($-250$ and $-840$~\kms, respectively). The low 1 and low 2 spectra of \ion{C}{4} are enlarged by a factor of 1.8 for the sake of clarity. The low 1 and low 2 spectra are shifted by $-0.5$ in all panels. Note the similarity between the absorption-line profiles of the low 1 and low 2 states.}
\end{center}
\end{figure}

\clearpage
\begin{figure}
\begin{center}
\includegraphics[angle=-90,width=13cm]{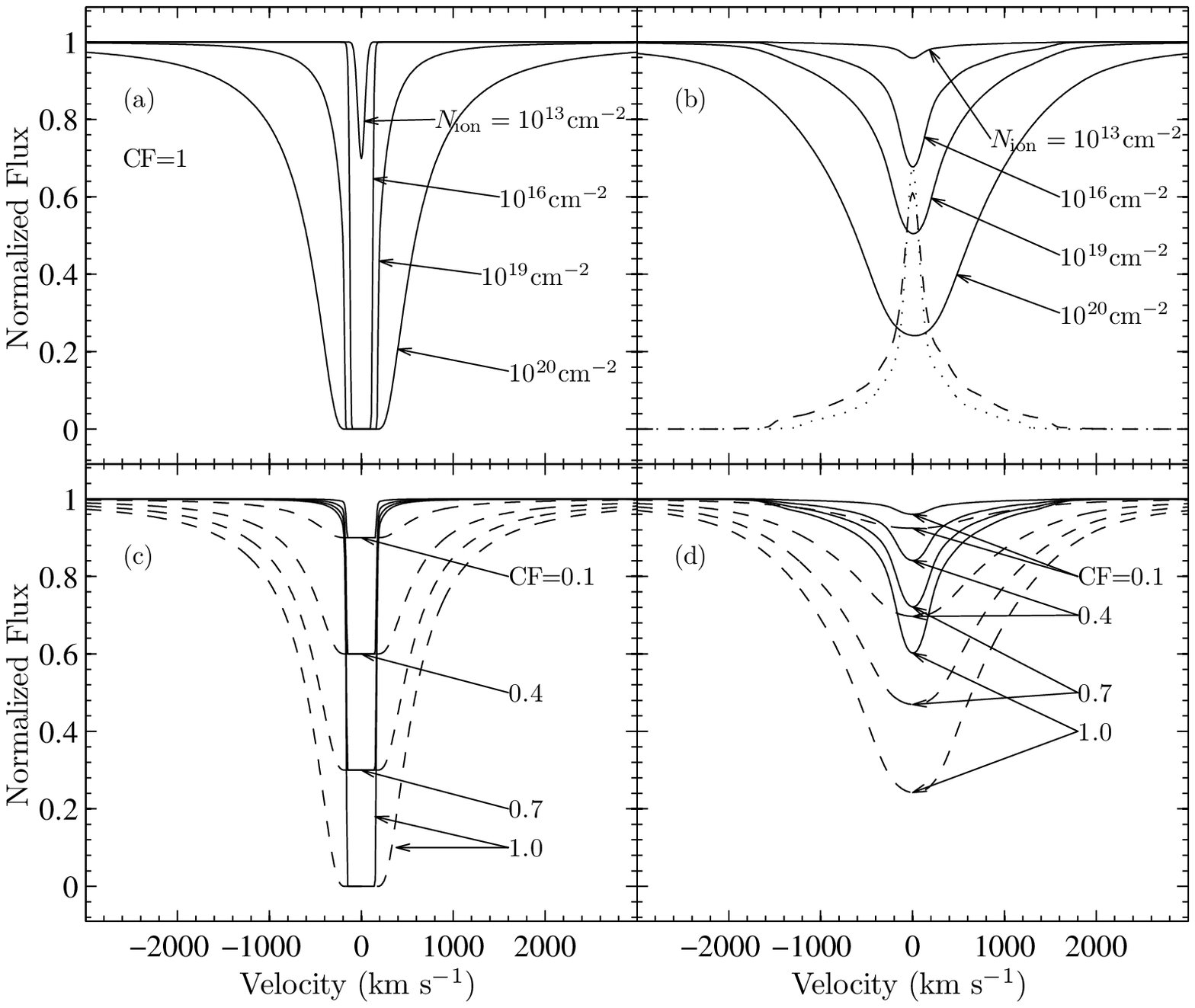}
\caption{The effect of instrumental broadening on the apparent absorption profile. The intrinsic absorption-line profiles (left panels) and the profiles after a convolution with the line-spread functions of \hst\ STIS (right panels) are plotted. A synthetic absorption line was used with $\lambda_{\rm ion}=1200$~\AA, $f_{\rm os}=1$ and $\Gamma=2.5\times10^8$~s$^{-1}$ (a typical $\Gamma$-value of the UV absorption lines analyzed in this paper). Top panels: absorption systems with $b=50$~\kms, CF=1, and a range of \nion\ (indicated). Note the broadening of the intrinsic absorption profiles as a result of the convolution with the instrumental LSF. The line-spread functions of the instrument at 1200~\AA\ (dashed) and 1500~\AA\ (dotted) are also plotted [panel (b)]. Bottom panels: absorption systems with $b=50$~\kms, \nion\ of $10^{18}$ and $10^{20}$~cm$^{-2}$ (solid and dashed lines, respectively), and a range of CFs (indicated). Note the significant weakening of the absorption depth as a result of the convolution with the instrumental LSF, and that it is practically impossible to differentiate a range of \nion\ at a given CF, from a range of CF at a given \nion\ [panels (b) and (d)].}
\end{center}
\end{figure}

\clearpage
\begin{figure}
\begin{center}
\includegraphics[angle=-90,width=12.7cm]{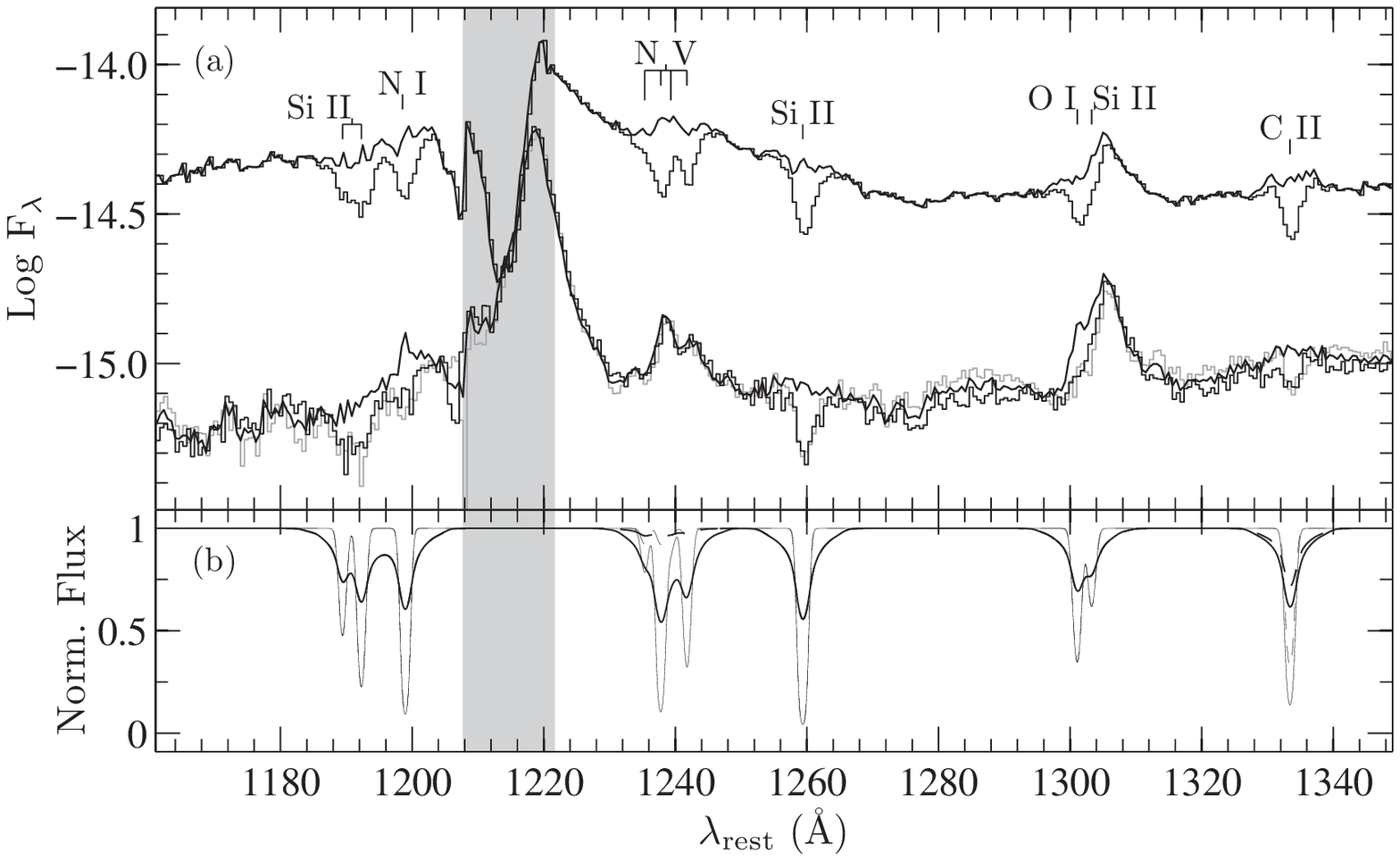}
\includegraphics[angle=-90,width=12.7cm]{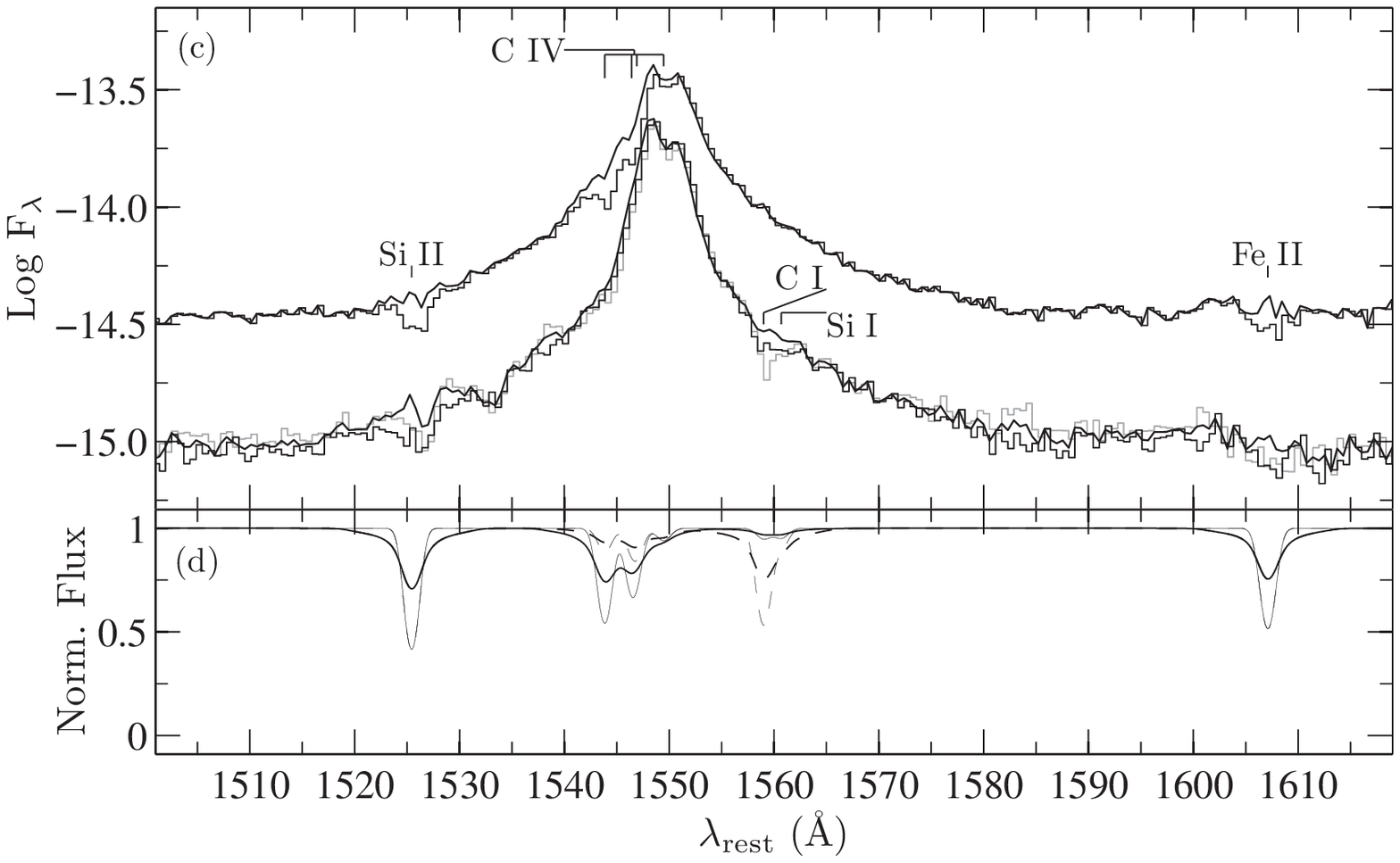}
\caption{The reconstructed spectra assuming an optically-thin foreground absorber with CF=1. Low 1, low 2, and high flux-states are plotted at the top panel of each plate using the same convention as in Fig.\ 1. The modeled intrinsic absorption-line profiles (thin line) and absorption-line profiles after a convolution with the instrumental LSF (thick line) are plotted in the bottom panel of each plate for the low- and high-states (dashed and solid lines, respectively).}
\end{center}
\end{figure}

\clearpage
\begin{figure}
\begin{center}
\includegraphics[angle=-90,width=12.7cm]{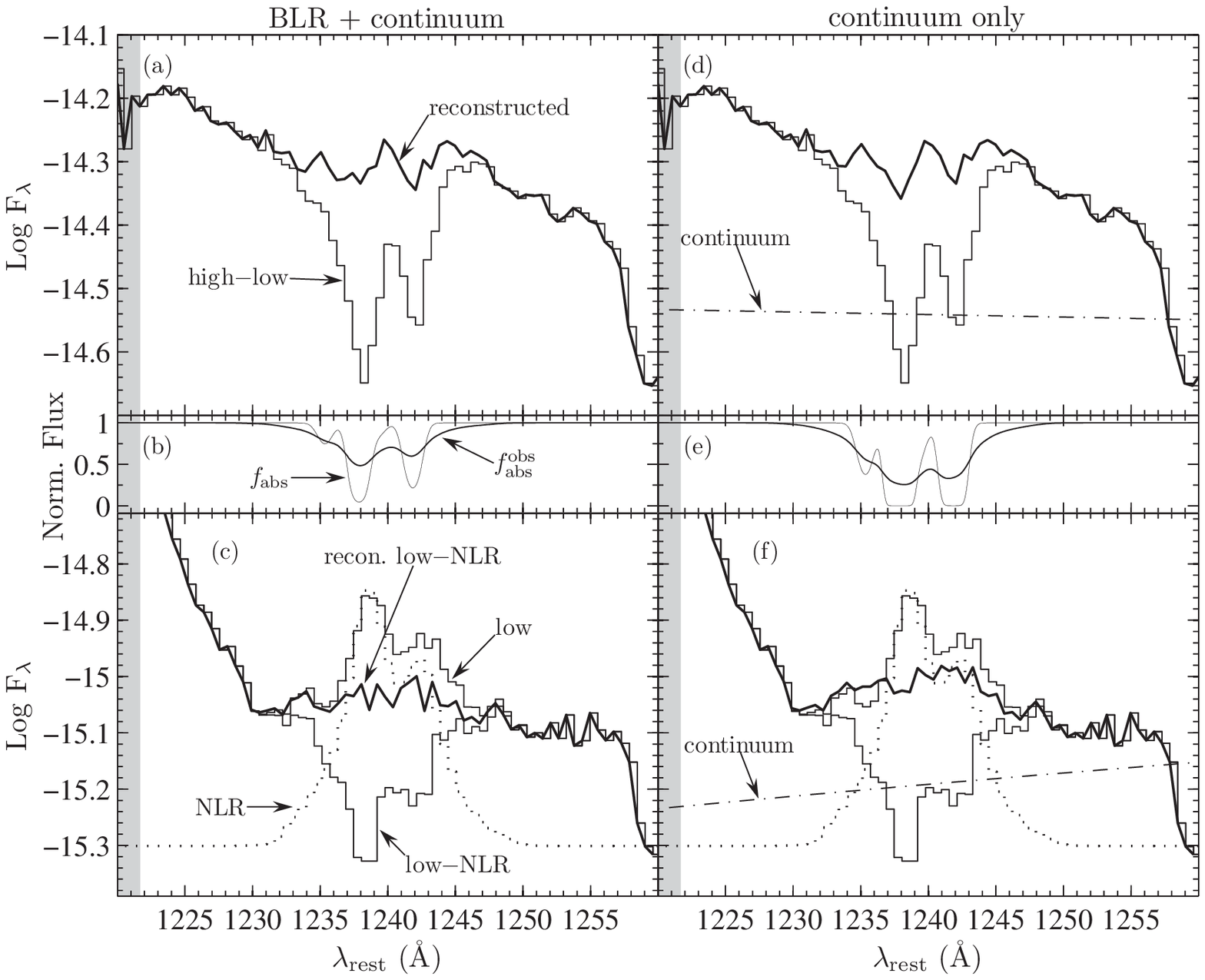}
\caption{The reconstruction of the \ion{N}{5} feature assuming an absorber located between the BLR and the NLR (left panels), or between the BLR and the continuum source (right panels). A constant-profile absorber is assumed for the low-state spectrum reconstruction. An optically-thin gas with CF=1 is assumed for both systems \Ah\ and B. Top panels: the difference spectrum between the high- and low-states and its reconstruction are presented (thin and thick lines, respectively), when in panel (d) the estimated continuum is also presented (dot-dashed). Middle panels: the modeled intrinsic absorption-line profiles (thin line) and absorption-line profiles after a convolution with the instrumental LSF (thick line). Bottom panels: the low spectrum (thin), the estimated \ion{N}{5} \ion{He}{2}-like narrow emission (dotted, shifted upwards for presentation purposes), their difference (i.e.\ the intrinsic BLR plus continuum spectrum; thin) and the reconstruction of the intrinsic BLR plus continuum spectrum (thick) are presented. In panel (f) the estimated continuum is also presented (dot-dashed). Note that the intrinsic BLR spectrum is reconstructed using the same absorption-line profiles measured for the high$-$low spectrum.}
\end{center}
\end{figure}

\clearpage
\begin{figure}
\begin{center}
\includegraphics[angle=-90,width=8cm]{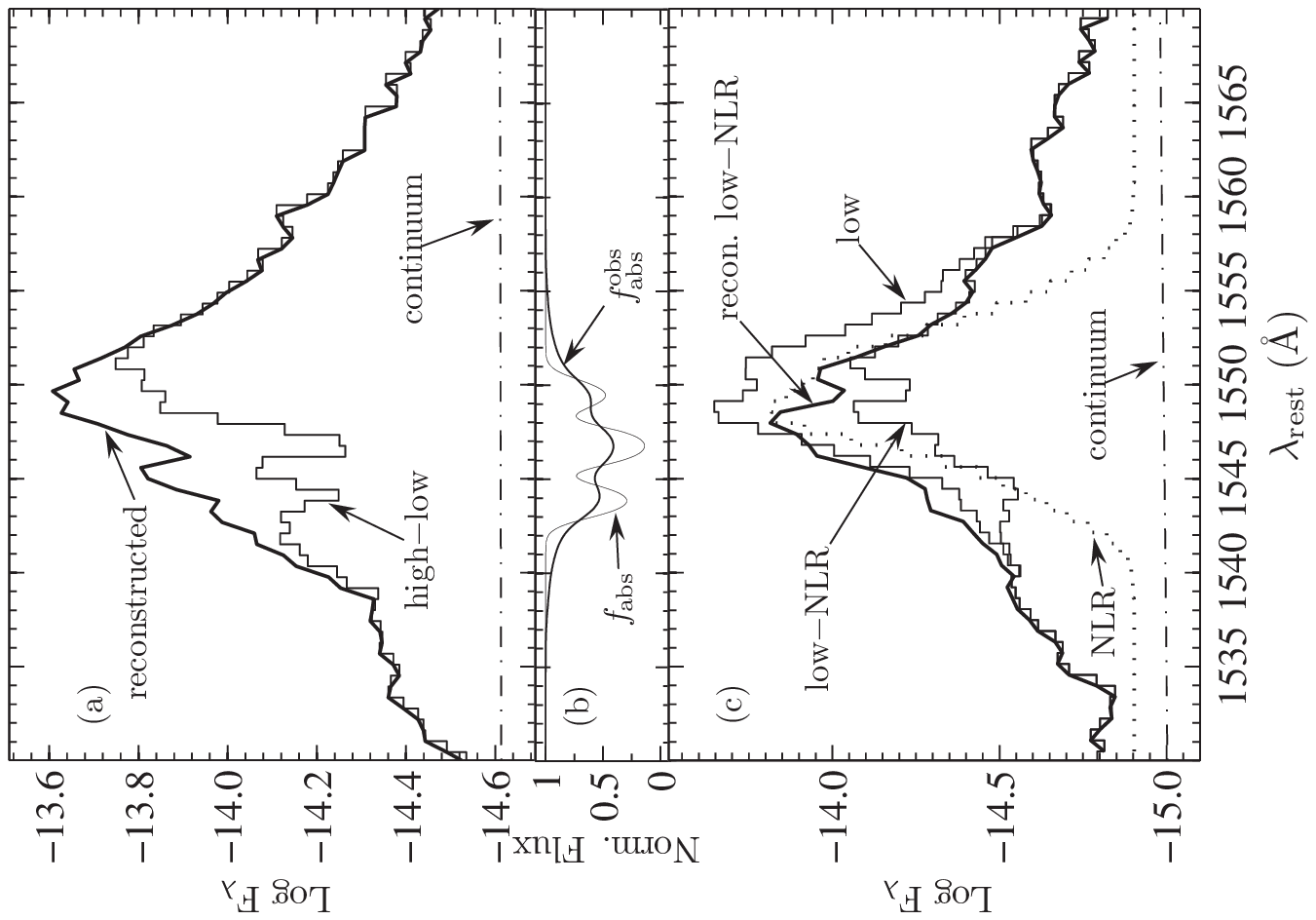}
\caption{Same as in Fig.\ 6 for the \ion{C}{4} feature only, assuming an absorber located between the BLR and the NLR. A constant-profile absorber is assumed for the low-state spectrum reconstruction. The estimated continua are plotted in panels (a) and (c). A reconstruction with an absorber between the BLR and the continuum source is not possible as the continuum level is too low to produce the observed absorption depth.}
\end{center}
\end{figure}


\begin{thebibliography}{}
\bibitem[\protect\citeauthoryear{Arav et~al.}{1999}]{1999ApJ...524..566A} Arav, N., Becker, R.\ H., Laurent-Muehleisen, S.\ A., Gregg, M.\ D., White, R.\ L., Brotherton, M.\ S., \& de~Kool, M.\ 1999, ApJ, 524, 566

\bibitem[\protect\citeauthoryear{Arav et~al.}{2002}]{2002ApJ...566..699A} Arav, N., Korista, K.\ T., \& de~Kool, M.\ 2002, ApJ, 566, 699

\bibitem[\protect\citeauthoryear{Baskin \& Laor}{2005}]{2005MNRAS...356..1029} Baskin, A., \& Laor, A.\ 2005, MNRAS, 356, 1029

\bibitem[\protect\citeauthoryear{Crenshaw et~al.}{1999}]{1999ApJ...516..750C} Crenshaw, D.\ M., Kraemer, S.\ B., Boggess, A., Maran, S.\ P., Mushotzky, R.\ F., \& Wu, C.-C.\ 1999, ApJ, 516, 750

\bibitem[\protect\citeauthoryear{Crenshaw et~al.}{2003}]{2003ARAA...41..117} Crenshaw, D.\ M., Kraemer, S.\ B., \& George, I.\ M.\ 2003, ARA\&A, 41, 117

\bibitem[\protect\citeauthoryear{Crenshaw et~al.}{2004}]{2004ApJ...612..152} Crenshaw, D.\ M., Kraemer, S.\ B., Gabel, J.\ R., Schmitt, H.\ R., Filippenko, A.\ V., Ho, L.\ C., Shields, J.\ C., \& Turner, T.\ J.\ 2004, ApJ, 612, 152

\bibitem[\protect\citeauthoryear{Desroches et~al.}{2006}]{2006ApJ...650...88D} Desroches, L.-B., et~al.\ 2006, ApJ, 650, 88

\bibitem[\protect\citeauthoryear{Ferland et~al.}{1998}]{1998PASP...110..761} Ferland, G.\ J., Korista, K.\ T., Verner, D.\ A., Ferguson, J.\ W., Kingdon, J.\ B., \& Verner, E.\ M.\ 1998, PASP, 110, 761

\bibitem[\protect\citeauthoryear{Filippenko \& Ho}{2003}]{2003ApJ...588L..13F} Filippenko, A.\ V., \& Ho, L.\ C.\ 2003, ApJ, 588, L13

\bibitem[\protect\citeauthoryear{Filippenko \& Sargent}{1989}]{1989ApJ...342L..11F} Filippenko, A.\ V., \& Sargent, W.\ L.\ W.\ 1989, ApJ, 342, L11

\bibitem[\protect\citeauthoryear{Filippenko, Ho \& Sargent}{1993}]{1993ApJ...410L..75F} Filippenko, A.\ V., Ho, L.\ C., \& Sargent, W.\ L.\ W.\ 1993, ApJ, 410, L75

\bibitem[\protect\citeauthoryear{Gabel et~al.}{2005a}]{2005ApJ...623...85G} Gabel, J.\ R., et~al.\ 2005a, ApJ, 623, 85

\bibitem[\protect\citeauthoryear{Gabel et~al.}{2005b}]{2005ApJ...631..741G} Gabel, J.\ R., et~al.\ 2005b, ApJ, 631, 741

\bibitem[\protect\citeauthoryear{Ganquly et~al.}{2001}]{2001ApJ...549..133G} Ganguly, R., Bond, N.\ A., Charlton, J.\ C., Eracleous, M., Brandt, W.\ N., \& Churchill, C.\ W.\ 2001, ApJ, 549, 133

\bibitem[Ganguly \& Brotherton(2008)]{2008ApJ...672..102G} Ganguly, R., \& Brotherton, M.~S.\ 2007, ApJ, 672, 102

\bibitem[\protect\citeauthoryear{George et~al.}{2000}]{2000ApJ...531...52G} George, I.\ M., Turner, T.\ J., Yaqoob, T., Netzer, H., Laor, A., Mushotzky, R.\ F., Nandra, K., \& Takahashi, T.\ 2000, ApJ, 531, 52

\bibitem[\protect\citeauthoryear{Hamann \& Ferland}{1999}]{1999ARA&A...37..487} Hamann, F., \& Ferland, G.\ 1999, ARA\&A, 37, 487

\bibitem[\protect\citeauthoryear{Hamann et al.}{2007}]{2007ASP...373..653} Hamann, F., Warner, C., Dietrich, M., \& Ferland, G.\ 2007, in ASP Conf.\ Ser.\ 373, The central engine of Active Galactic Nuclei, ed.\ L.\ C.\ Ho, \& J.-M.\ Wang (San Francisco: ASP), 653

\bibitem[\protect\citeauthoryear{Haynes et~al.}{1998}]{1998AJ...115..62} Haynes, M.\ P., Hogg, D.\ E., Maddalena, R.\ J., Roberts, M.\ S., \& van~Zee, L.\ 1998, AJ, 115, 62

\bibitem[\protect\citeauthoryear{Iwasawa et~al.}{2000}]{2000MNRAS...318..879} Iwasawa, K., Fabian, A.\ C., Almaini, O., Lira, P., Lawrence, A., Hayashida, K., \& Inoue, H.\ 2000, MNRAS, 318, 879

\bibitem[\protect\citeauthoryear{Kaspi et~al.}{2004}]{2004AJ....127.2631K} Kaspi, S., Brandt, W.\ N., Collinge, M.\ J., Elvis, M., \& Reynolds, C.\ S.\ 2004, AJ, 127, 2631

\bibitem[\protect\citeauthoryear{Kraemer et~al.}{1999}]{1999ApJ...520..564K} Kraemer, S.\ B., Ho, L.\ C., Crenshaw, D.\ M., Shields, J.\ C., \& Filippenko, A.\ V.\ 1999, ApJ, 520, 564

\bibitem[\protect\citeauthoryear{Kraemer et~al.}{2001}]{2001ApJ...551..671K} Kraemer, S.\ B., et~al.\ 2001, ApJ, 551, 671

\bibitem[\protect\citeauthoryear{Kraemer et~al.}{2006}]{2006ApJS..167..161K} Kraemer, S.\ B., et~al.\ 2006, ApJS, 167, 161

\bibitem[\protect\citeauthoryear{Laor}{2004}]{2004ASP...311..169} Laor, A.\ 2004, in ASP Conf.\ Ser.\ 311, AGN physics with the Sloan Digital Sky Survey, ed.\ G.\ T.\ Richards, \& P.\ B.\ Hall (San Francisco: ASP), 169

\bibitem[\protect\citeauthoryear{Laor}{2006}]{2006ApJ...643..112} Laor, A.\ 2006, ApJ, 643, 112

\bibitem[\protect\citeauthoryear{Laor}{2007}]{2007ASP...373..384} Laor, A.\ 2007, in ASP Conf.\ Ser.\ 373, The central engine of Active Galactic Nuclei, ed.\ L.\ C.\ Ho, \& J.-M.\ Wang (San Francisco: ASP), 384

\bibitem[\protect\citeauthoryear{Laor \& Brandt}{2002}]{2002ApJ...569..641L} Laor, A., \& Brandt, W.\ N.\ 2002, ApJ, 569, 641

\bibitem[\protect\citeauthoryear{Laor et~al.}{1997}]{1997ApJ...489..656P} Laor, A., Jannuzi, B.\ T., Green, R.\ F., \& Boroson, T.\ A.\ 1997, ApJ, 489, 656

\bibitem[\protect\citeauthoryear{Lira et~al.}{1999}]{1999MNRAS...305..109} Lira, P., Lawrence, A., O'Brien, P., Johnson, R.\ A., Terlevich, R., \& Bannister, N.\ 1999, MNRAS, 305, 109

\bibitem[\protect\citeauthoryear{McHardy et~al.}{2003}]{2003MNRAS...348..783} McHardy, I.\ M., Papadakis, I.\ E., Uttley, P., Page, M.\ J., \& Mason, K.\ O.\ 2003, MNRAS, 348, 783

\bibitem[\protect\citeauthoryear{Moran et~al.}{1999}]{1999PASP...111..801} Moran, E.\ C., Filippenko, A.\ V., Ho, L.\ C., Shields, J.\ C., Belloni, T., Comastri, A., Snowden, S.\ L., \& Sramek, R.\ A.\ 1999, PASP, 111, 801

\bibitem[\protect\citeauthoryear{Moran et~al.}{2005}]{2005AJ...129..2108} Moran, E.\ C., Eracleous, M., Leighly, K.\ M., Chartas, G., Filippenko, A.\ V., Ho, L.\ C., \& Blanco, P.\ R.\ 2005, AJ, 129, 2108

\bibitem[\protect\citeauthoryear{Morton}{1991}]{1991ApJS...77..119M} Morton, D.\ C.\ 1991, ApJS, 77, 119

\bibitem[\protect\citeauthoryear{Murphy et~al.}{1996}]{1996ApJS...105..369} Murphy, E.\ M., Lockman, F.\ J., Laor, A., \& Elvis, M.\ 1996, ApJS, 105, 369

\bibitem[\protect\citeauthoryear{Netzer}{1997}]{1997Ap&SS...248..127} Netzer, H.\ 1997, Ap\&SS, 248, 127

\bibitem[\protect\citeauthoryear{Netzer \& Laor}{1993}]{1993ApJ...404L..51N} Netzer, H., \& Laor, A.\ 1993, ApJ, 404, L51

\bibitem[\protect\citeauthoryear{Ogle et~al.}{2004}]{2004ApJ...606..151} Ogle, P.\ M., Mason, K.\ O., Page, M.\ J., Salvi, N.\ J., Cordova, F.\ A., McHardy I.\ M., \& Priedhorsky W.\ C.\ 2004, ApJ, 606, 151

\bibitem[\protect\citeauthoryear{O'Neill et~al.}{2006}]{2006ApJ...645..160O} O'Neill, P.\ M., et~al.\ 2006, ApJ, 645, 160

\bibitem[Osterbrock]{1989agna.book.....O} Osterbrock, D.\ E.\ 1989, Astrophysics of Gaseous Nebulae and Active Galactic Nuclei (Mill Valley, CA: University Science Books)

\bibitem[\protect\citeauthoryear{Peterson et~al.}{2004}]{2004ApJ...613..682P} Peterson, B.\ M., et~al.\ 2004, ApJ, 613, 682

\bibitem[\protect\citeauthoryear{Peterson et~al.}{2005}]{2005ApJ...632..799P} Peterson, B.\ M., et~al.\ 2005, ApJ, 632, 799. Erratum, 2006, ApJ, 641, 638

\bibitem[\protect\citeauthoryear{Pradhan \& Peng}{1995}]{1995aelm.conf....8P} Pradhan. A.\ K., \& Peng, J.\ 1995, in Space Telescope Science Institute Symposium Series No. 8, ed.\ R.\ E.\ Williams, \& M.\ Livio (Cambridge University Press)

\bibitem[\protect\citeauthoryear{Roy et~al.}{1996}]{1996ApJ...460..284} Roy, J.-R., Belly, J., Dutil, Y., \& Martin, P.\ 1996, ApJ, 460, 284

\bibitem[Rybicki \& Lightman(1979)]{1979rpa..book.....R} Rybicki, G.\ B., \& Lightman, A.\ P.\ 1979, Radiative Processes in Astrophysics (New York: Wiley)

\bibitem[\protect\citeauthoryear{Savage et~al.}{2000}]{2000ApJS...129...563} Savage, B.\ D., et~al.\ 2000, ApJS, 129, 563

\bibitem[Schlegel, Finkbeiner, \& Davis(1998)]{1998ApJ...500..525S} Schlegel, D.\ J., Finkbeiner, D.\ P., \& Davis, M.\ 1998, ApJ, 500, 525

\bibitem[\protect\citeauthoryear{Seaton}{1979}]{1979MNRAS...187..73P} Seaton, M.\ J.\ 1979, MNRAS, 187, 73P

\bibitem[\protect\citeauthoryear{Shemmer et~al.}{2003}]{2003MNRAS...343..1341} Shemmer, O., Uttley, P., Netzer, H., \& McHardy, I.\ M.\ 2003, MNRAS, 343, 1341

\bibitem[Shih, Iwasawa \& Fabian(2003)]{2003MNRAS...341..973} Shih, D.\ C., Iwasawa, K., \& Fabian, A.\ C.\ 2003, MNRAS, 341, 973

\bibitem[\protect\citeauthoryear{Springob et~al.}{2005}]{2005ApJS...160..149} Springob, C.\ M., Haynes, M.\ P., Giovanelli, R., \& Kent, B.\ R.\ 2005, ApJS, 160, 149

\bibitem[\protect\citeauthoryear{Thim et~al.}{2004}]{2004AJ...127..2322} Thim, F., Hoessel, J.\ G., Saha, A., Claver, J., Dolphin, A., \& Tammann, G.\ A.\ 2004, AJ, 127, 2322

\bibitem[\protect\citeauthoryear{Uttley et~al.}{2000}]{2000MNRAS...312..880} Uttley, P., McHardy, I.\ M., Papadakis, I.\ E., Cagnoni, I., \& Fruscione, A.\ 2000, MNRAS, 312, 880

\bibitem[\protect\citeauthoryear{Vaughan et~al.}{2005}]{2005MNRAS...356..524} Vaughan, S., Iwasawa, K., Fabian, A.\ C., \& Hayashida, K.\ 2005, MNRAS, 356, 524

\bibitem[\protect\citeauthoryear{Vestergaard}{2003}]{2003ApJ...599..116V} Vestergaard, M.\ 2003, ApJ, 599, 116

\bibitem[\protect\citeauthoryear{Wells}{1999}]{1999JQSRT..62..29} Wells, R.\ J.\ 1999, JQSRT, 62, 29
\end{thebibliography}
\end{document}